\documentclass{aastex}          
\usepackage{spr-astr-addons}    
\usepackage{url}

\begin{document}
\title{Star Formation in the starburst cluster in NGC~3603}
   
\slugcomment{Based on early release science observations with the WFC3 and on
observations taken at the ESO Very Large Telescope within the observing program
082.C-0889(A)}

\shorttitle{Star formation in NGC~3603}
\shortauthors{M. Correnti et al.}

\author{Matteo Correnti}
\affil{INAF - Istituto di Astrofisica Spaziale e Fisica Cosmica, Via P. Gobetti, 
101, I-40129 Bologna, Italy}	           
\email{correnti@iasfbo.inaf.it}

\author{Francesco Paresce}
\affil{INAF - Istituto di Astrofisica Spaziale e Fisica Cosmica, Via P. Gobetti, 
101, I-40129 Bologna, Italy}	           

\author{Rossella Aversa}		   
\affil{INAF - Istituto di Astrofisica Spaziale e Fisica Cosmica, Via P. Gobetti,
101, I-40129 Bologna, Italy}	           

\author{Giacomo Beccari}		   
\affil{ESO - European Southern Observatory, Karl-Schwarzschild-Str. 2, D-85748 
Garching bei M\"unchen, Germany}		   

\author{Guido De Marchi}
\affil{ESA, Space Science Department, Keplerlaan 1, 2200 AG Noordwijk,
The Netherlands}   
	
\author{Marcella Di Criscienzo}
\affil{INAF - Osservatorio Astronomico di Roma, Via Frascati 33, Monte Porzio 
Catone, Roma, Italy}
	   
\author{Xiaoying Pang}
\affil{Astronomisches Rechen-Institut, Heidelberg University, Heidelberg, 
Germany}
		   
\author{Loredana Spezzi}
\affil{ESA, Space Science Department, Keplerlaan 1, 2200 AG Noordwijk,
The Netherlands}   

\author{Elena Valenti} 
\affil{ESO - European Southern Observatory, Karl-Schwarzschild-Str. 2, D-85748 
Garching bei M\"unchen, Germany}		   

\and 

\author{Paolo Ventura} 
\affil{INAF - Osservatorio Astronomico di Roma, Via Frascati 33, Monte Porzio 
Catone, Roma, Italy}
	  
\date{Received; accepted}

\begin{abstract}
We have used new, deep, visible and near infrared observations of the compact
starburst cluster in the giant HII region NGC~3603 and its surroundings with the
WFC3 on HST and HAWK-I on the VLT to study in detail the physical properties of
its intermediate mass ($\sim$ 1 - 3 $M_\odot$) stellar population. We show that
after correction for differential extinction and actively accreting stars, and
the study of field star contamination, strong evidence remains for a continuous
spread in the ages of pre-main sequence stars in the range $\sim 2$ to $\sim 30$
Myr within the temporal resolution available. Existing differences among
presently available theoretical models account for the largest possible
variation in shape of the measured age histograms within these limits. We also
find that this isochronal age spread in the near infrared and visible
Colour-Magnitude Diagrams cannot be reproduced by any other presently known
source of astrophysical or instrumental scatter that could mimic the luminosity
spread seen in our observations except, possibly, episodic accretion. The
measured age spread and the stellar spatial distribution in the cluster are
consistent with the hypothesis that star formation started at least 20-30 Myrs
ago progressing slowly but continuously up to at least a few million years ago.
All the stars in the considered mass range are distributed in a flattened oblate
spheroidal pattern with the major axis oriented in an approximate South-East -
North-West direction, and with the length of the equatorial axis decreasing with
increasing age. This asymmetry is most likely due to the fact that star
formation occurred along a filament of gas and dust in the natal molecular cloud
oriented locally in this direction. 
\end{abstract}


\section{Introduction}
\label{Intro}

Hertzsprung-Russell (HR) or Colour-Magnitude Diagrams (CMDs) in combination with
theoretical models can be used, in principle, to determine the age distribution
of stars in a cluster. Measurable luminosity spreads in these diagrams have been
found in many, if not all, young nearby and massive clusters measured to date.
They can be interpreted as real age spreads provided other sources of luminosity
scatter can be defined accurately and their effects taken properly into account.
These sources include, but may not be limited to, observational errors,
differential extinction and reddening, distance uncertainties, field star
contamination, unresolved multiplicity, variability and position shifts in the
CMD due to the accretion process. The difficulties associated with identifying
and quantifying this long list of possible non age dependent contributors to the
luminosity spread has seriously hampered work in this field and created
skepticism or ambiguity as to the reality of any claimed significant age spread
in these clusters \citep{hartmann03,hillenbrand09,jeffries11}.\\
\begin{figure*}
\centering
\includegraphics[width=16cm]{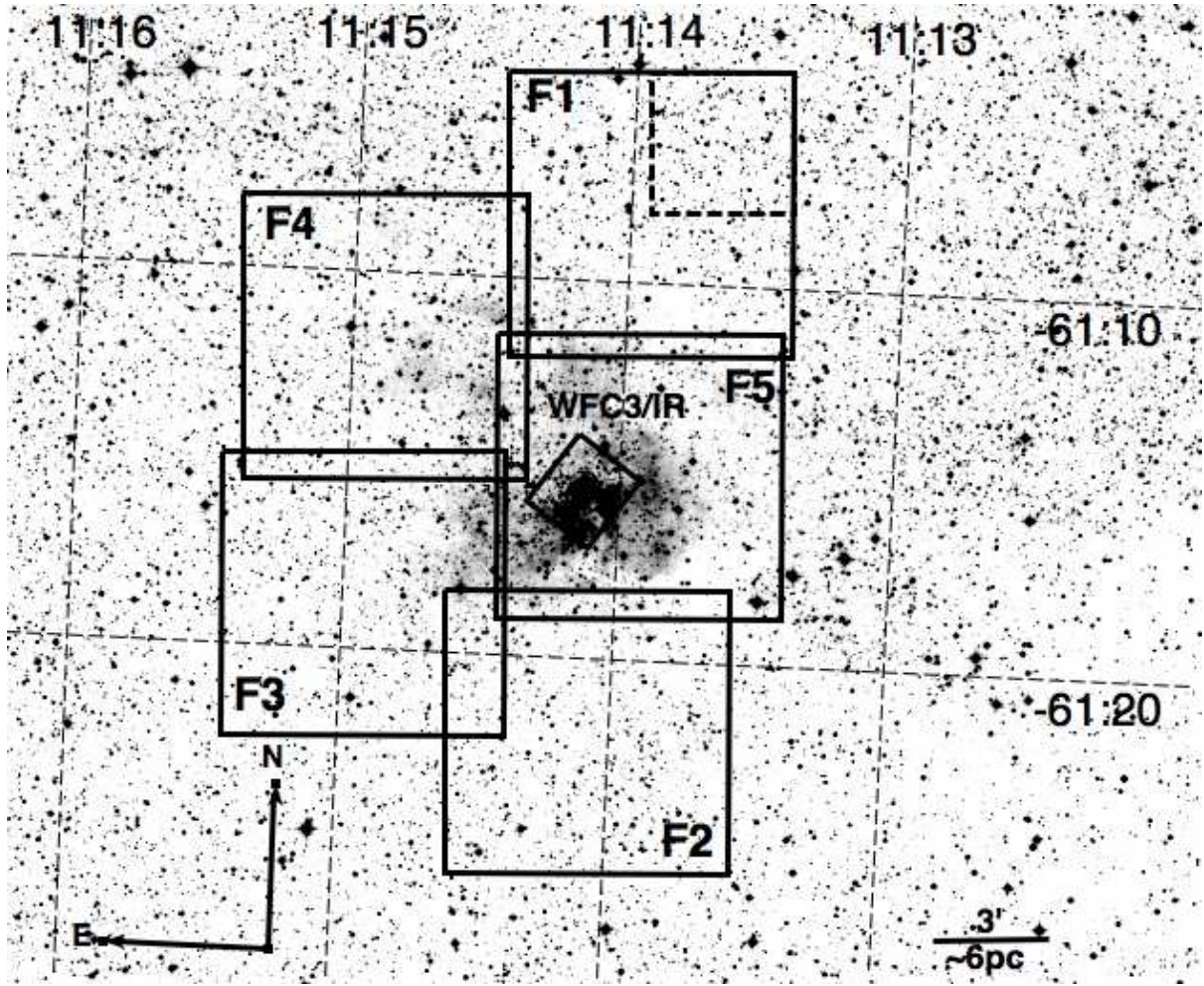}
\caption{WFC3 (small square) and HAWK-I (F1-F5) FoV plotted on the red
Digitalized Sky Survey plates. Image centred at $RA(J2000) = 11^h 15^m 7^s.26$
and $Dec(J2000) = -61^{\circ} 15' 37".9$, as determined by SB04. Orientantion
and size scale are also shown in the image. The position of frame 3 of
field F1, used to derive the right hand CMD of Fig.\ref{cmdraw}, is shown as a
dashed square.}
\label{figFoV}
\end{figure*}
In the specific case of the starburst cluster of radius $2.5'$ located in the
giant HII region NGC~3603 \citep[][hereafter NPG02]{nurn02b,nurn02} at a
distance from the Sun of $6.0 \pm 0.8$ kpc \citep[see][for a complete
discussion]{harayama08}, the question of its age and possible spread has been
hotly debated for some time. This cluster is one of the most compact and
luminous massive star clusters in the Milky Way (MW), with a bolometric
luminosity of 100 times that of the Orion Cluster, and a possible nearby
prototype of the extragalactic starburst clusters like NGC~2070 in the Large
Magellanic Cloud (LMC). \citet{melnick89} were the first to suggest on the basis
of theoretical isochrones on a reasonably deep $V,B-V$ CMD, the existence of a
small ($\pm 2$ Myr) real age spread around an average age of 2.5 Myr in mainly
relatively massive Main Sequence (MS) stars with star formation seemingly
propagating roughly from North to South. Some time later, on the other hand,
\citet{eisenhauer98} using near infrared (IR) $J$, $H$, $K$ ground based
observations argued that their data were consistent with a single burst of star
formation and that the cluster was actually a coeval group of stars of age 1-2
Myr with no need for any significant age spread. This conclusion essentially was
echoed, with some variations, by subsequent investigators \citep[][hereafter
SB04]{stolte04,sung04}.\\  More recently, \citet{melena08}, although supporting
these later conclusions for the most massive stars ($> 100 M_\odot$) in the
core, argue that lower mass objects ($20-40 M_\odot$) may have a larger spread
of 1-4 Myr.  \citet[][hereafter HEM08]{harayama08}, also find for the lower mass
Pre-Main Sequence (PMS) stars an average age of 0.7 Myr with a possible spread
of only $\pm 0.3$ Myr, but that for the heavier MS stars an upper age limit of
2.5 Myr is appropriate.  \citet{rochau10} instead show that most intermediate
mass PMS stars in the core of the cluster have an age of 1 Myr but with a sparse
population of low mass stars of age about 4 Myr.\\ Finally, \citet[][hereafter
B10]{beccari10}, working in the visible, using a much larger sample of stars and
exploiting the technique of separating PMS cluster members from field objects on
or near the Zero Age Main Sequence (ZAMS) using the H$\alpha$ line in emission
\citep{demarchi10} conclude that the single age or starburst hypothesis is not
consistent with their data. Instead they indicate that the PMS cluster members
distribute themselves essentially continuously across a wide range of ages
between 1 and 30-40 Myr with a peak at a few Myr. This conclusion could, in
principle, be negatively affected by the possible errors in isochronal ages
discussed above.\\ In any case, this large uncertainty in the correct age
attribution of the cluster stars is a very important issue to resolve as it
affects the crucial question of how and where star formation occurs in the
progenitor Giant Molecular Cloud (GMC) of which the compact starburst cluster is
only a small part. In particular, one can ask whether star formation in these
clusters occurs in a slow ($\sim 10-20$ Myr) or fast ($\sim 1$ Myr) mode and if
it is localized or distributed over a large volume of the cloud. In this
context, an important question left unanswered is whether or not the older
H$\alpha$ emitting stars detected by B10 actually belong to the starburst
cluster or whether they are part of a more spatially distributed population
belonging to the much larger GMC surrounding it, much like the situation, for
example, of the cluster Tr~14 and Tr~15 in the Carina nebula
\citep{preibisch11}.\\ In order to tackle these issues, we have analyzed deep
near IR Wide Field Camera 3 (WFC3) and High Acuity Wide field K-band Imager
(HAWK-I) images of the starburst cluster in NGC~3603 (from now on simply
referred to as NGC~3603). The WFC3 IR and HAWK-I datasets represent a
significant improvement in accuracy in a wavelength region where high instrument
sensitivity and lower interstellar gas extinction should lead to a much better
determination and understanding of the variables involved in establishing the
sources of the luminosity spread observed in the visible.  In particular, the IR
is best suited to minimize the effects of colour and luminosity shifts due to
the accretion process in these PMS stars. In the visible CMD, the Ultraviolet
(UV) excess emission in the $V$ band from this process results in younger stars
looking older and, thus, possibly biasing the putative age spread. In the IR
CMD, on the other hand, the IR excess works in the opposite direction making
older stars look younger with the result that the oldest population cannot be
contaminated by even older objects that don't exist. In addition, these two
datasets allow us to sample the entire cluster together with its surroundings
well into the vast majority of its stellar population. In this paper, we will
consider and attempt to quantify the possible measurable sources of scatter in
the HR diagram of the cluster, investigating evidence for and against age
spread, and we will try to demonstrate that the scatter is most likely due to a
real age effect and not to the ``nuisance'' sources listed above.\\ The plan of
the paper is the following. In Sect.~2 we present the observations of NGC~3603
obtained with WFC3 and HAWK-I, while in Sect.~3 we describe the data reduction
processes used to derive the sources' final catalogue. In Sect.~4 we present the
analysis done on the CMDs with the histograms representing the number of PMS
stars in the cluster in each age interval, as a function of age and our
derivation of the structure and the spatial distribution of the cluster. In
Sect.~5, we analyze the sources of uncertainty that can bias our work and we
discuss how the results can be affected. Finally, we summarize and discuss our
results in  Sect.~6.

\section{Observations}
\label{Obs}

The photometric data used in this work consist of a series of deep multi-band
images acquired with the WFC3 on board the Hubble Space Telescope (HST) and 
HAWK-I on the Very Large Telescope (VLT). Observations obtained with these two
instruments are described in the next subsections.

\subsection{WFC3 observations}
\label{WFC3obs}

The WFC3 consists of two detectors, one optimized for observations in the
wavelength range 200 to 1000 {\it nm} (UVIS channel) and the other between 0.9
and 1.7 $\mu$m (IR channel). The UVIS detector consists of $2\times2K\times4K$
thinned, backside illuminated, UV optimized e2v CCDs covering a field of view
(FoV) of $162'' \times 162''$ at a plate scale of $0.04''$/px. The IR detector
is a $1K \times 1K$ Teledyne HgCdTe FPA, MBE grown, substrate removed detector
offering a total FoV of $123'' \times 136''$ at a pixel resolution of $0.13''$.
A more detailed description of the WFC3 and its current performance can be found
in \citet{mackenty10} and \citet{wong10}.\\ The WFC3 data used in this work are
part of the Early Release Science (ERS) observations obtained by the WFC3
Scientific Oversight Committee for the study of star forming regions in nearby
galaxies (Program ID number 11360). NGC~3603 was observed using both the UVIS
and IR channels. The UVIS and IR datasets were presented in B10 and in
\citet[][hereafter S11]{spezzi11},  respectively. Briefly, in the IR, NGC~3603
was observed through the F110W ({\it J} band) and F160W ({\it H} band) broadband
filters, the F127M, F139M and F153M medium-band filters, and the F128N narrow
band filter. Three images were acquired for each band in order to allow for the
removal of cosmic rays, hot pixels, and other detector blemishes. Details on the
total exposures for each band are reported in Table~2 of S11. The position of
the WFC3 field is shown in Fig.~\ref{figFoV}. The WFC3  $J$ band image of the
cluster has been presented in Fig.~1 of S11 and a five colour composite image
can be found in the Hubble Heritage site:
\url{http://heritage.stsci.edu/2010/22/index.html}.\\ 

\subsection{HAWK-I observations}
\label{Hawkiobs}

In addition to the WFC3 data, we used a series of observations obtained with
HAWK-I, in order to sample the cluster stellar population beyond the central
region in the IR. HAWK-I \citep[see][]{kissler08} is a near-IR imager at the ESO
8m (VLT-UT4, Yepun) equipped with a mosaic of four Hawaii 2RG $2048 \times 2048$
pixel detectors with a scale of $0.106''$ per pixel. The camera has a total FoV
on the sky of $7.5' \times 7.5'$ with a small cross-shaped gap of $\sim 15''$
between the four detectors. The observations of NGC~3603 were performed in March
2009, in visitor mode, and retrieved from the ESO archive (Proposal ID:
082.C-0889(A), PI N\"urnberger). The observing conditions were generally good,
with an average seeing measured on image of $\sim 0.6''$. The data are derived
from images taken through the standard broadband $J, H$, and $K_s$ filters.\\ A
total of five fields were observed, one roughly centred on the cluster and four
fields sampling the external regions around the cluster core (see
Fig.~\ref{figFoV}) covering a total area of about 337 $arcmin^2$. Each frame
acquired through the $H$ and $K_s$ filters is the combination of 6 exposures
each 10 sec long, while frames taken with the $J$ filter are the result of 2 and
1 exposures each 30 and 60 sec long, respectively. For each filter and pointing
the observation was repeated with a random dithering pattern and jitter box
width of $30''$ until reaching a total exposure time of 10 min for each filter.
As can be seen in Fig.~\ref{figFoV}, each field has a region of overlap with the
adjacent ones, so that it is possible to check the accuracy of the astrometric
and photometric calibrations.

\section{Data reduction}
\label{Data}

The photometry of the entire WFC3 dataset was performed on the flat-fielded
(FLT) images following a standard Point Spread Function (PSF) fitting procedure
with DAOPHOT II/ALLSTAR \citep{stetson87} package. An accurate PSF model for
each image was obtained adopting a first-order polynomial on about 150 isolated
and well exposed stars homogeneously distributed in the FoV. A master list of
stars was obtained using stars detected in the F110W image (the deepest of the
IR dataset) and was then used as input for ALLFRAME  \citep{stetson94} to
perform an accurate PSF fitting on all the images. All the magnitude values for
each star were normalized to a reference frame and averaged together, and the
photometric errors were derived as the standard deviation of the repeated
measurements. The magnitudes were finally transformed into the VEGAMAG
photometric system by adopting the recipe described in  \citet[][,see S11 for
details]{kalirai09}. The number of objects detected at least in one of the five
filters is 9693.\\ The final WFC3 catalogue, obtained  imposing that stars have
been measured in each filters (i.e.: F110W, F160W,  F127M, F128N, F139M and
F153M), contains 8831 stars. Saturation of the WFC3 images occurs at $J \sim 12$
and $H \sim 11.5$, while stars are detected at $3\sigma$ accuracy down to
$J=20.5$ and $H=19.5$, respectively. At the distance and typical
age of NGC~3603, these limiting magnitudes correspond roughly to cluster members
with masses in the range between 0.5 and 3.5 $M_\odot$, according to the models
of \citet[][hereafer DC09]{dicrisci09}.\\ Photometry in the core region, out to
a distance of $\sim 10''$ from the cluster centre derived by SB04 and adopted
also in B10 is severely affected by strong camera saturation due to the high
concentration of very bright young massive O-B stars. We decided to excise this
region from our analysis, so the final catalogue does not include stars inside
it, due to the fact that the completeness in this part is very low (see
Sect.~\ref{Completeness} for the discussion about completeness in the WFC3 and
HAWK-I fields) and the photometric uncertainty is very high in the detected
stars. After this cut, the final WFC3 catalogue contains 8553 stars.\\  For what
concerns the HAWK-I data, we used standard IRAF routines to process them. For
all filters, we derived a sky image from a median combination of the dithered
images that we subtracted from each frame. To normalize the pixel-to-pixel
response, all frames were divided by a normalized twilight flat. Finally, all
the flat- and sky-corrected frames have been averaged in a single image for each
of the three filters. In what follows, we refer to a frame as a combination of
these sets. Aperture photometry was performed independently on each frame using
the aperture photometry code SExtractor \citep{bertin96} and  adopting a fixed
aperture radius of 4 pixels ($0.4''$). The lists of magnitudes for each filter
were then combined with the requirement that each star had to be measured at the
same time in the $J,H$ and $K_s$ bands. The mosaic of the five pointings samples
a total of 110347 stars from the magnitude $J \sim 12$ down to $J \sim 22$ in an
area of $\sim 15' \times 22.5'$ around the cluster centre. We estimate that
objects as faint as $J \sim 21$, $H \sim 20$ and $K_s \sim 20$ are detected at a
$S/N \geq 10$.\\ More than 10000 stars from the Two Micron All Sky Survey
\citep[2MASS,][]{skrutskie06} catalogue, from a total sample of $\sim 31000$
2MASS sources in the area of the HAWK-I fields, were used as photometric and
astrometric standards to obtain accurate photometric calibration of the $J, H$
and $K_s$ bands, and to transform the instrumental relative position of stars
into J2000 celestial coordinates. The HAWK-I catalogue sampling the cluster
central regions was adopted as catalogue of secondary astrometric standards in
order to properly find an astrometric solution for the WFC3 catalogue. We
estimate that the global uncertainty in the astrometric solution relative to
2MASS is $\sim 0.2''$. Magnitudes $J, H$ and $K_s$ of all the frames have been
calibrated using the stars in common with 2MASS, typically $\sim 200$ stars for
each of them, and the results of the calibration have been checked comparing the
magnitude values for the stars in common between different fields, i.e.: the
stars in the overlapping region. Each overlapping strip contains $\sim 500$
stars, except in the case of field F2 and F5, where the stars in common are
$\sim 300$. The differences in the independent calibration for the magnitudes
$J, H$ and $K_s$ are of order $\sigma_j \sim 0.04$, $\sigma_H \sim 0.04$ and
$\sigma_K \sim 0.05$, respectively reassuring us about the precision of the
calibration through 2MASS.\\ To compare magnitudes derived from the WFC3 and
from HAWK-I, and to build a single catalogue containing all the sources in the
FoV, it is necessary that one of the magnitudes system is translated into the
other, due to the different filters between WFC3 and HAWK-I. The $J$ and $H$
(F110W and F160W) magnitudes in the WFC3 system have been transformed to the
HAWK-I (2MASS) system, using empirical transformations, derived from 3308 stars
in common between the two fields. In particular, we obtained the {\it new} $J$
and $H$ adopting the following equations:
\begin{eqnarray}                 
J = F110W - 0.111 - 0.246 * (F110W - F160W)\\
H = F160W - 0.049 - 0.175 * (F110W - F160W)
\end{eqnarray}
The final total catalogue, obtained combining WFC3 and HAWK-I fields, with the
obvious prescription of not double counting the stars in the overlap regions,
contains 112265 stars. For those stars having both a WFC3 and HAWK-I
measurement, only the more precise 2MASS calibrated WFC3 magnitude is retained
in the final catalogue.

\subsection{Completeness of WFC3 and HAWK-I fields}
\label{Completeness}

\begin{figure}
\plotone{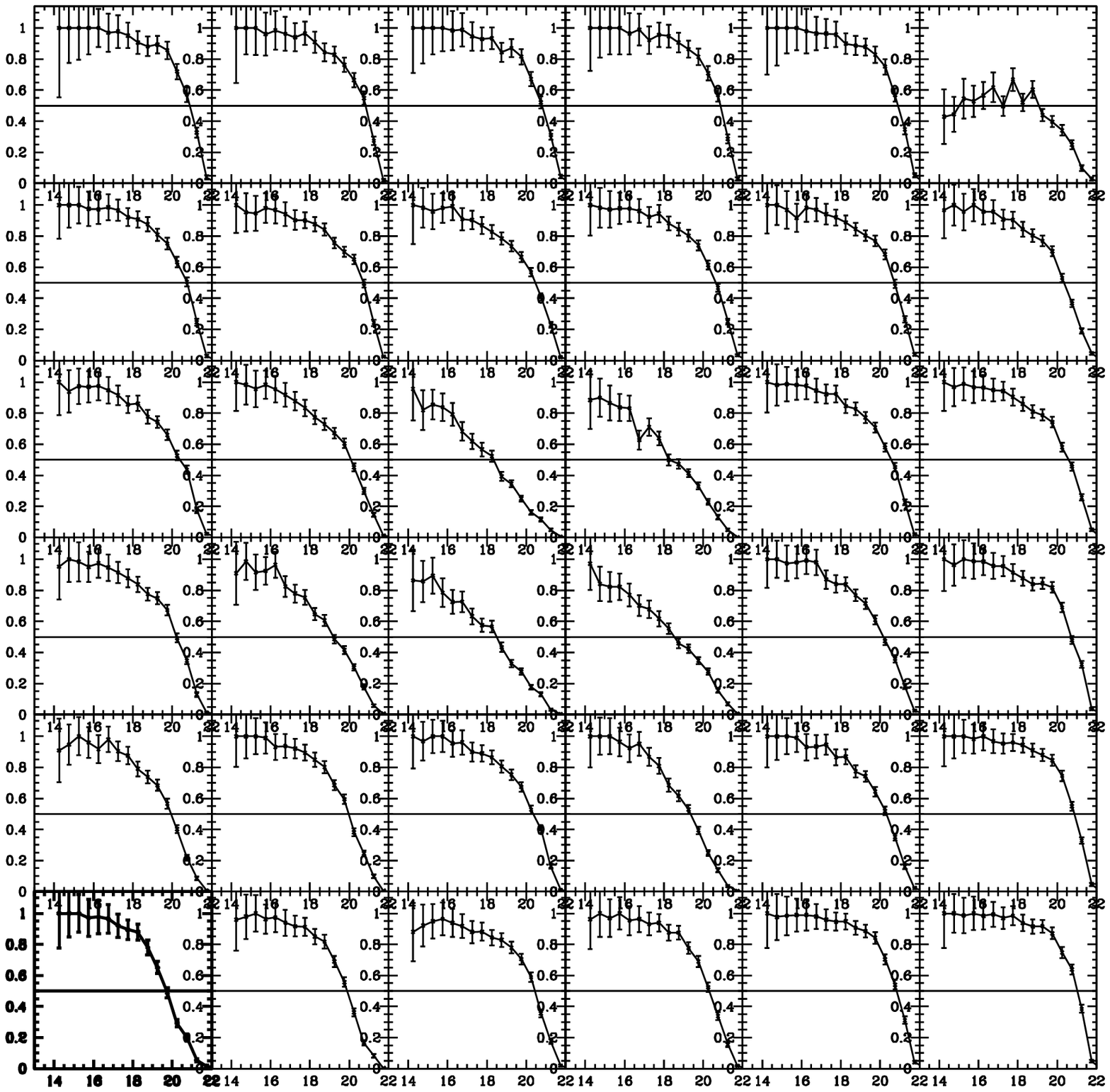}
\plotone{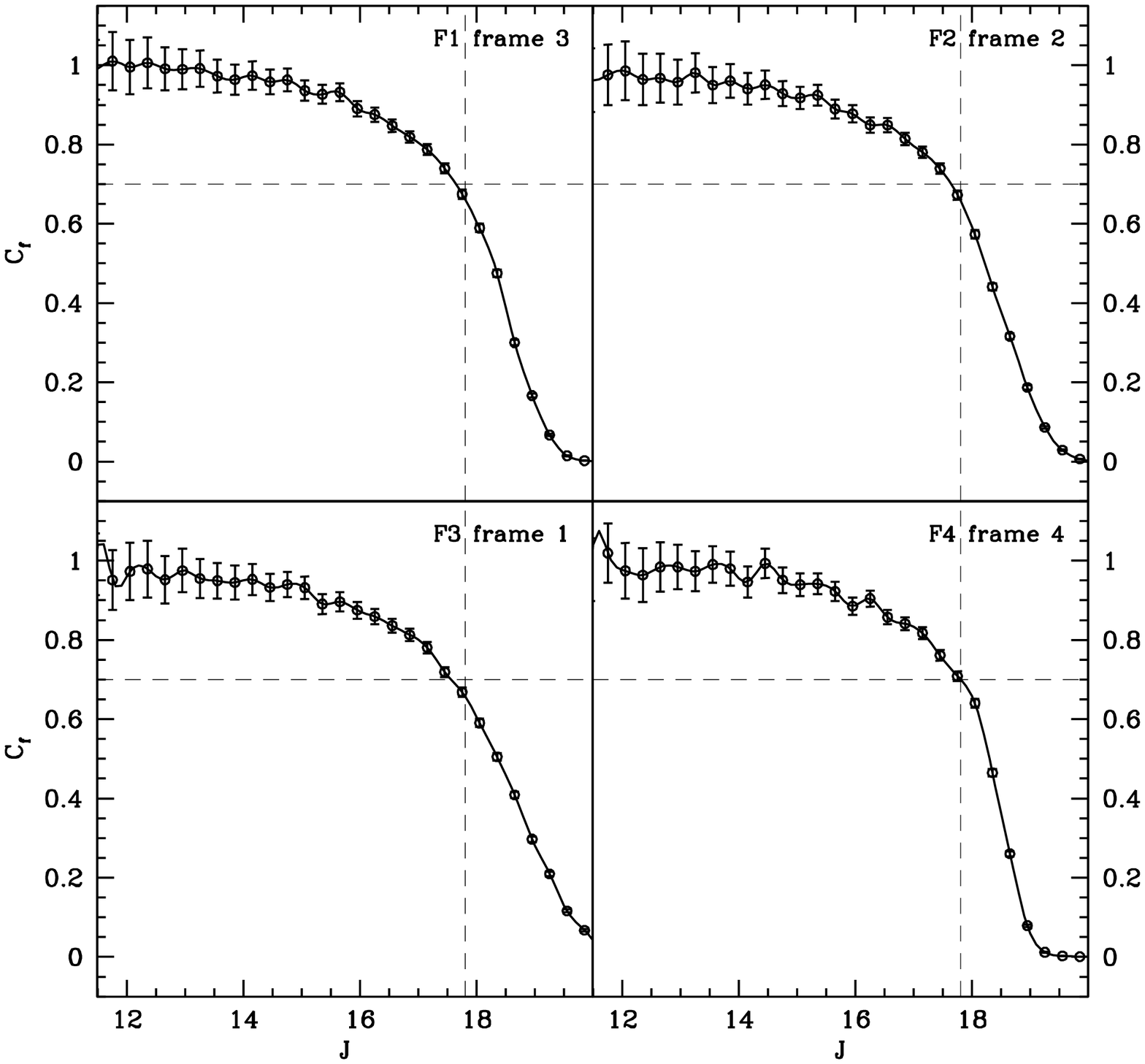}
\caption{Top panel: completeness factor as a function of $J$ magnitude for
different sub-regions in the WFC3 fov. The reported lines are best-fitting
curves to the data. Completeness does not show variations with position and/or
radial distance from the centre; the four central panel show the low  $C_f$ in
the excised central region (i.e.: region inside a radius of 10'' from cluster
centre). Bottom panel: completeness factor as a function of $J$ magnitude for
the four analyzed frames of the four different HAWK-I fields. Also in this case,
the reported lines are best-fitting curves to the data. The completeness factor
is similar in all the four analyzed frames. Our limiting magnitude ($J =
17.83$)and a  $C_f = 0.7 (i.e.: 70\%)$ are also indicated (dashed vertical and
horizontal  lines, respectively).}  
\label{cf}
\end{figure}

As the last step of the reduction process, we calculated the completeness in the
WFC3 field and in four HAWK-I fields, from F1 to F4, selecting a different frame
in each field, in order to avoid a possible bias. We do not calculate the
completeness in each frame because the crowding conditions in the observed
external fields are not critical and are very homogeneous in general. This
suggests that the completeness level should be not very different from frame to
frame. On the contrary, since the WFC3 field is very crowded and, as we have
mentioned above, the central region is affected by strong camera saturation, due
to the high concentration of very bright young massive O-B stars, the situation
is more complex.\\ We performed extensive artificial star experiments in one
frame for each field assumed to be representative of the whole field sample
following the recipe described in \citet{bellazzini02}. A
total of nearly 150000 artificial test stars were added to each selected field.
The artificial stars where uniformly distributed in each run into grids of
cells so that the minimum distance between two artificial star is of the order
of $\sim5$ Full Width Half Maximum (FWHM), in order to leave the crowding
conditions unaltered. The test stars were distributed in magnitude according to
a luminosity function similar to the observed one but monotonically increasing
also beyond the limit of the photometry and with a colour distribution covering
the full colour ranges of the observed stars. The entire reduction procedure has
been repeated in the same way on each synthetically enriched image. We derive
that the completeness factors for the four  HAWK-I frames are the same within
the errors in the range of magnitude analyzed ($12 \lesssim  J \lesssim 18$).
This is not unexpected since the crowding conditions are nearly similar, and in
any case very far from the critical conditions found in the central region of
the cluster. Furthermore, the spatial distribution of stars is very homogeneous
so there is no expected variation of completeness with position. This result
provides robust support for taking the derived completeness as representative of
the whole HAWK-I fields, except the F5, since it has been demonstrated that the
observed field-to-field variations have minimal impact on the overall
completeness. Concerning the WFC3, the completeness factor, $C_f$, is
$\geq 85\%$ for $J < 18$, in the external regions, while falls to lower values
in the central ones due to crowding and saturation (see top panels of
Fig.~\ref{cf}). As stated  before, in the HAWK-I fields instead, the behaviour
of $C_f$ is more linear and we derive a $C_f > 70\%$ for $J < 18$ in all the
four analyzed frames (see bottom panels of Fig.~\ref{cf}). 
The completeness situation for field F5 is slightly different due its dependence
on distance from the centre. In the external regions of this field, the 
completeness is similar to what we found for the other HAWK-I fields, i.e.  $C_f
> 70\%$ for $J < 18$, while it decreases in the crowded central portion of the
field, in the region of overlap with the WFC3 field. However, it is worth 
noting that in this specific region, as mentioned at the end of the previous
section (Sec.~\ref{Data}), we retain in the final catalogue only stars from the
WFC3 catalogue, so the reference completeness is that of the WFC3 shown in the
top panels of Fig.~\ref{cf}. These panels show that there is no variation with
position and/or radial distance from the centre, so the combination of these two
effects makes it unnecessary to consider the completeness variation encountered
in the field F5, and to consider it in the same way as the other HAWK-I
fields.  

\begin{figure}
\plotone{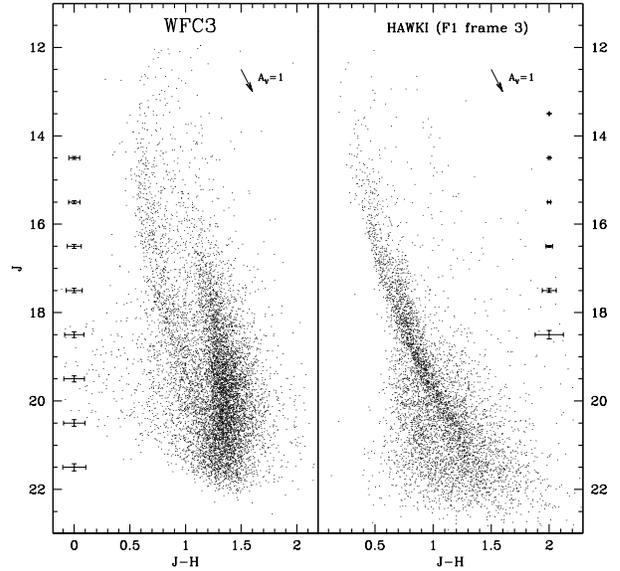}
\caption{Observed $J,J-H$ CMDs of the WFC3 field (left panel), representative of
the central region of the cluster, and of the frame 3 of HAWK-I field F1 (right
panel), which position is reported in Fig.\ref{figFoV}, sampling the external
regions. Both CMDs have been obtained using fields of the same area. Photometric
colour and magnitude uncertainties are reported in both CMDs. For the HAWK-I
one, we obtained them from completeness simulations. (comparison between the
input magnitudes and the output ones). The reddening vector ($A_V$=1) is also
shown.}  
\label{cmdraw}
\end{figure}

\subsection{CMDs of WFC3 and HAWK-I fields}
\label{secCMD}

The result of the reduction process described above is shown in the form of two
observed $J,J-H$ CMDs in Fig.~\ref{cmdraw}. In the left panel, we have the CMD
of the WFC3 field, representative of the central region of the cluster, while in
the right panel we have the CMD of  a frame of the F1 HAWK-I field (F1\_3, see
for a visual reference Fig.~\ref{figFoV}), well outside the nominal $2.5'$
radius of the cluster (NPG02).  The HAWK-I field is chosen such that the areas
are the same for a meaningful comparison.\\ Photometric colour and magnitude
uncertainties are reported in both CMDs. The errors for the WFC3 CMD were
derived as the simple standard deviations of the repeated measurements of the
WFC3 field magnitudes, while photometric errors for the HAWK-I field have been
derived from the mean values of the $mag_{in}-mag_{out}$ distribution, derived
from the incompleteness simulations. In the WFC3 CMD, two distinct and well
separated features are clearly visible: on the blue side, the almost vertical
sequence around $J-H \sim 0.6-0.7$ bending to the red at $J \sim 18$, is
consistent with a suitably reddened ZAMS of field and cluster stars (see also
Sec.~\ref{profile} and \ref{dens_map}). On the red side, the wide clump of stars
around $J-H \sim 1.3-1.4$ less evident at brighter magnitudes, but well
populated from $J \sim 17.0$ to $J \sim 19.5$ is consistent with the position of
cluster PMS stars. In support of this conjecture, in the right panel, the HAWK-I
CMD clearly shows only the corresponding ZAMS of the field stars with no sign of
the PMS clump. Hence, the different number of stars and the different shape of
the two CMDs are not due to crowding, photometric uncertainties or completeness
but to the different population of stars we are observing. 

\section{Analysis}
\subsection{Reddening Correction}
\label{Redd_corr}

The data discussed above were dereddened using different approaches for the two
datasets. B10 and S11 obtained dereddened magnitudes of the stars in the WFC3
FoV by exploiting the extinction study of NGC~3603 by SB04. These authors found
that the value $A_V = 4.5$ is representative of the very centre of the cluster
while there is an increase toward the external regions. Adopting the
total-to-selective extinction ratio ($R_V$) value of 3.55, suggested by SB04,
B10 estimated that the value $A_V = 5.5$ is representative of the mean visual
extinction in the area sampled by the WFC3 observations. S11 adopted this value
of $A_V$ and the extinction law of \citet{cardelli89} to derive
the following extinction for $J$ and $H$ magnitudes: $A_J \sim 2.00$ and $A_H
\sim 1.05$, respectively. Following S11, we adopted these values to correct the
magnitudes of stars in our catalogue for the WFC3 field.\\ Concerning the
HAWK-I fields, instead, we adopted the approach suggested by HEM08, namely that
in the observed CMDs for each field, we determine the position of the mean ridge line
of the ZAMS that is the predominant feature present in the external CMDs as seen
for example in the right panel of Fig.~\ref{cmdraw}. As a theoretical template,
we adopt the ZAMS by DC09. The idea is that, if there is a discrepancy in the
two ZAMS (observed vs theoretical), the only free parameter that allows us to
shift the mean ridge line in the colour-magnitude plane representing the
observed ZAMS is the extinction $A_V$. Therefore we used a $\chi^2$ test to
derive the $A_V$ value for which the difference in colour and magnitude between
the mean ridge line representing the observed ZAMS and the theoretical ZAMS is
minimum. This process has been applied to each frame of HAWK-I fields so that we
can obtain a frame-to-frame reddening correction. We observed that fields on the
northern side (F1 and some frames of F4) have a lower extinction respect to the
central field and, above all, resepect to the southern ones.\\ In detail, we
have that the mean value of extinction, $<A_V>$, for each field is the
following: $A_V = 4.3 \pm 0.2$ for F1, $A_V = 5.1\pm 0.3$ for F2, $A_V = 4.9 \pm
0,4$ for F3, $A_V = 5.0 \pm 0.4$ for F4 and finally $A_V = 4.9 \pm 0.4$ for F5.
Thus it is clear that the choice of a single value of $A_V$ for all the HAWK-I
fields would lead to an overestimate of the extinction in the majority of frames
and more in general in the observed fields.\\ In conclusion, the magnitudes of
stars in the final catalogue have been corrected in the appropriate manner,
depending on their position. All the results in the following, if not explicitly
said, have been obtained using this reddening correction obtained with the
prescription described above. 

\begin{figure*}
\centering
\hspace{-0.8cm}
\plotone{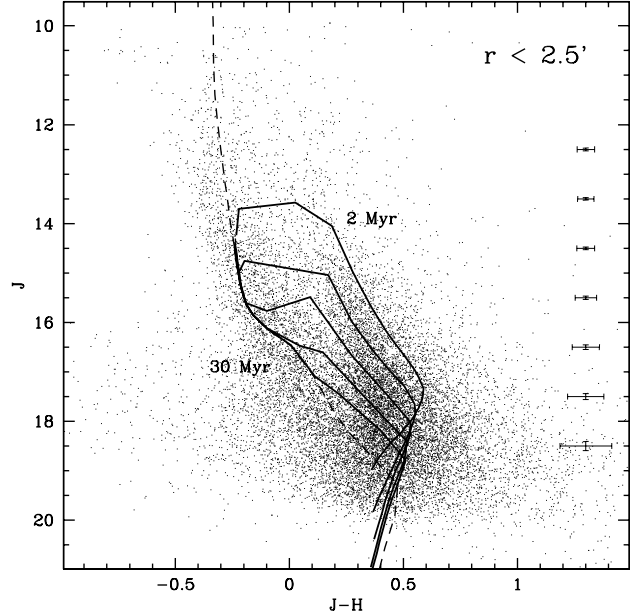}
\caption{Dereddened $J,J-H$ CMD using stars located inside the nominal radius
$r =2.5'$ (from NPG02) of the cluster. The dashed line is the ZAMS from DC09,
for solar metallicity and adopting a distance modulus $(m-M)_0=13.90$. The
population of PMS stars is fitted with $2,5,10,20,30$ Myr PMS isochrones (from
right to left), also from DC09 (solid lines). The position of the 2 and 30 Myr
isochrones is indicated. On the right side the photometric colour and magnitude
uncertainty are reported.}
\label{cmdiso}
\end{figure*}

\subsection{Theoretical models and age histograms}
\label{iso_hist}
 
\begin{figure}
\plotone{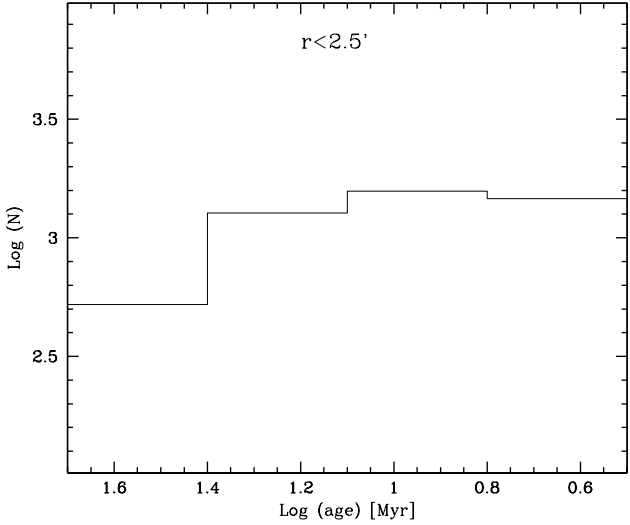}
\caption{Logarithmic histogram representing the number of PMS stars in the
cluster with magnitude range $13.6 < J < 17.8$, in each age interval, as a 
function of age. The histogram is derived using stars  located inside the
nominal radius $r=2.5'$ (NPG02) and isochrones as in Fig.~\ref{cmdiso}} 
\label{hist}
\end{figure}

The dereddened data were used to determine the age distribution of PMS stars in
the cluster. In Fig.~\ref{cmdiso}, we show the dereddened
$J_0,(J-H)_0$\footnote{For brevity, in the following, if not explicitly stated,
all the reported magnitudes and colours are reddening-corrected (i.e., for
example, $J=J_0=$ extinction corrected $J$ magnitude)} CMD of the entire
cluster, namely a region with a radius of $2.5'$ from the cluster centre,
derived exploiting the WFC3+HAWK-I catalogue. We use here the radius for the
cluster given by NPG02. The position of the ZAMS is taken from DC09 for solar
metallicity (dashed line), having adopted a distance modulus $(m-M)_0 = 13.90$
(see HEM08 and references therein). Together with the ZAMS we also plot the PMS
theoretical isochrones of DC09, for ages $2, 5, 10, 20, 30$ Myr. This set of
isochrones is characterized by a non grey atmosphere and $\alpha = 1.5$, where
$\alpha$ is the parameter representing the convection efficency (see
Sect.~\ref{diff_model} and the Appendix \ref{Appendix} for a detailed
description of the prescriptions and processes necessary to obtain them).
Stars located above the 2 Myr isochrone, i.e. $J \sim 13$ and $J-H \sim 0.55$,
are mainly stars younger than 1-2 Myr. In addition, a small fraction of these
are probably stars having a different reddening respect to the mean value
adopted in this work so that they have been corrected with a wrong value of
reddening and they would need in principle to be shifted in the CMD.  
Note that the photometric uncertainty of our data (see crosses in
Fig.~\ref{cmdiso}) confirms that it is possible to assign relative ages to these
stars with an accuracy of a factor of 2.\\ In principle, age histograms are an
excellent tool to determine the age distribution of stars in clusters. To obtain
them, we need to count objects in each age interval limited by the position of
the isochrones in CMDs. The process is not trivial in the IR because, as can be
seen in Fig.~\ref{cmdiso}, isochrones are not always parallel to each other, in
contrast with what happens in the visible (see for example Fig.~3 of B10). In
the IR, they start to cross at $J \sim 17.8$, so we must exclude stars fainter
than this limit, considering only brighter sources in our analysis.  In fact,
without this magnitude cut, older PMS stars, on the blue side of the CMD, would
be artificially more numerous than younger ones. On the other hand, the
brightest magnitude we can consider is reached at $J \sim 13.6$ by the 2 Myr
isochrone.\\ Thus, the actual magnitude range, in which we can perform our
counts, is $13.6 < J < 17.8$ corresponding roughly to a mass range from 1 to 3
$M_\odot$. This sample corresponds to $\simeq 30\%$ ($\sim 5000$) of the total
number of the stars detected in the cluster. We used as a lower age limit the
isochrone of 2 Myr because younger isochrones cross the older ones at brighter
magnitudes (the 1 Myr isochrone crosses the 2 Myr one at $J \sim 17$), thus
reducing the range of analysis. Moreover, we selected this lower limit also
because at lower magnitudes (i.e. $J \sim 18-19$) the completeness factor $C_f$
is much lower and the photometric uncertainties on the $J-H$ colour are
comparable, if not bigger, than the typical isochrone separation; so our
conservative choice eliminates the possibility of systematic errors in the age
determination of stars. We also decided to use the ZAMS as the blue colour limit
since, as we will see clearly in the density profile and maps shown in
Sect.~\ref{dens_map}, a good fraction of old ($\sim 20-30$ Myr) cluster PMS
stars are actually located near the ZAMS and would be excluded from our sample
using a redder limit. The corresponding age distribution of the stars in
Fig.~\ref{cmdiso} is shown in the form of a logarithmic histogram in
Fig.~\ref{hist}. This figure clearly shows that star formation in NGC~3603 has
been ongoing for at least 20-30 Myr and no visible gaps are present, or at least
evident, with the level of age resolution we adopted (a logarithmic bin of 0.3,
corresponding to a factor of 2 in age).

\subsection{Comparison between UVIS and IR histograms}
\label{histUV_IR}

\begin{figure}
\plotone{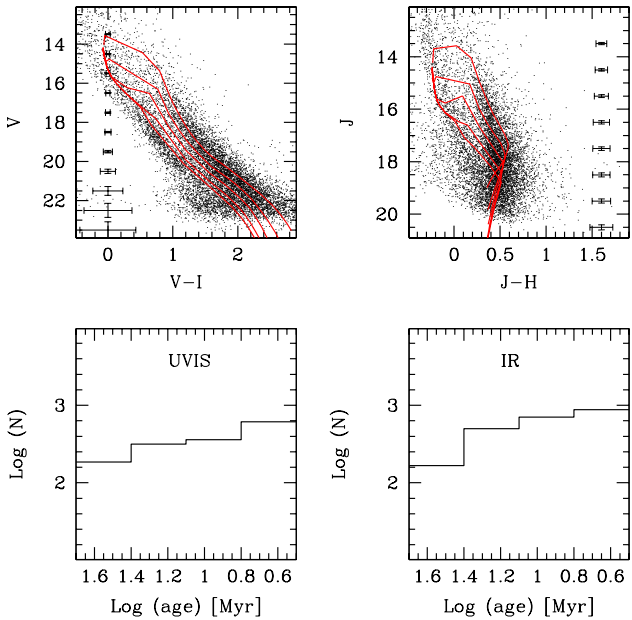}
\caption{Top panels: Dereddened $V,V-I$ (left panel) and $J,J-H$ (right 
panel) CMD of the WFC3 FoV. The population of PMS stars is fitted with
$2,5,10,20,30$ Myr PMS isochrones (from right to left) from DC09 (red solid
lines).Photometric colour and magnitude uncertainty are reported for both CMDs.
Bottom panels: logarithmic histograms representing the number of PMS stars in
the cluster, in each age interval, as function of age, in UV and IR FoV. The
histograms are derived from the isochrones showed in top panels.}
\label{uvir}
\end{figure}

To test the results obtained with the DC09 theoretical models in the IR, we
compared them with the histogram obtained in the UV with the same models. In the
top panels of Fig.~\ref{uvir}, we show the $V,V-I$ and $J,J-H$ CMDs in the WFC3
FoV, with, superimposed, 2, 5, 10, 20, 30 Myr isochrones, from DC09. As already
said, we observe that in the UV CMD the isochrones are parallel to each other
and so it is possible to extend the analysis to redder colours (i.e.: younger
stars), while in the IR CMD we are limited by the shape of the isochrones that
at a given magnitude (depending on the model) start to cross each other. To have
a direct comparison between the resulting histograms, therefore, we excluded the
age range from 1 to 2 Myr given by the UV theoretical models and we limited
the analysis to the magnitude range $16 < V < 20$. The bottom panels
of Fig.~\ref{uvir} show the logarithmic histograms representing the age
distribution of stars in the two cases. In both histograms, the age spread is
present and the two distributions are quite similar in shape within the adopted
age uncertainty of a factor of two.  

\subsection{Structure of the cluster}
\label{profile}

\begin{figure}
\plotone{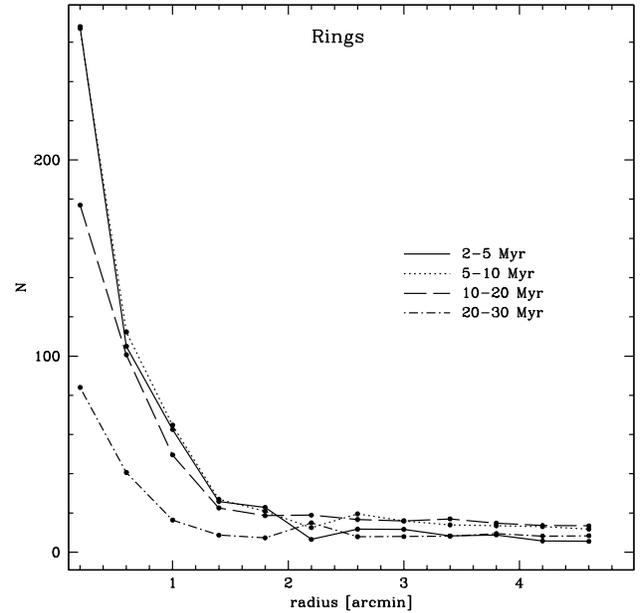}
\caption{Number of sources, per unit area, as a function of distance from the
centre of the cluster. Solid line: stars with age between 2 and 5 Myr; dotted
line: stars with age between 5 and 10 Myr; long dashed line: stars with age
between 10 and 20 Myr; dotted-dashed line: stars with age between 20 and 30
Myr. All objects are mostly concentrated within $r=2.5'$.}
\label{profTOT}
\end{figure}

\begin{figure}
\plotone{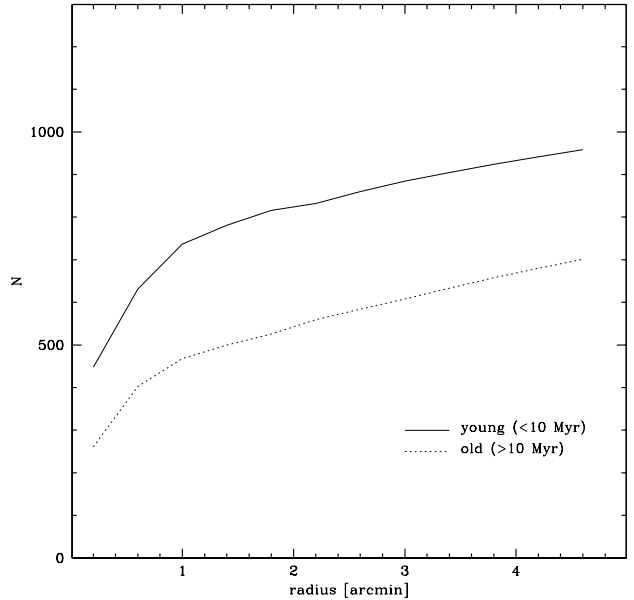}
\caption{Cumulative radial profiles obtained from young stars in the age
interval between 2 and 10 Myr (solid line) and from old stars in the age
interval between 10 and 30 Myr (dotted line). The number of young stars is
substantially twice respect to the old ones.  Both distributions have the same
shape, with a steep increase until $r \sim 2.5'$ and a flattening beyond it.}
\label{cumPROF}
\end{figure}

\begin{figure}
\plotone{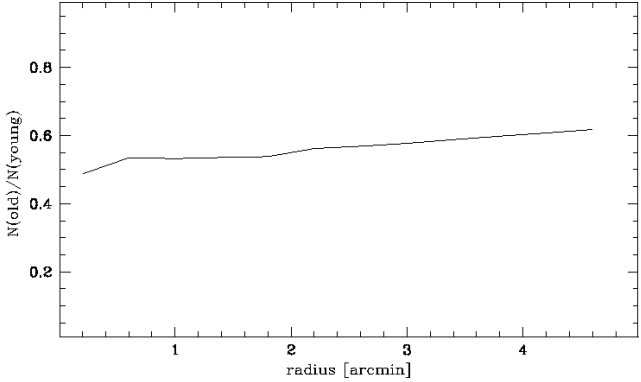}
\caption{Old/young ratio derived from the cumulative distributions shown in
Fig.~\ref{cumPROF}.}
\label{cumRATIO}
\end{figure}

\begin{figure}
\plotone{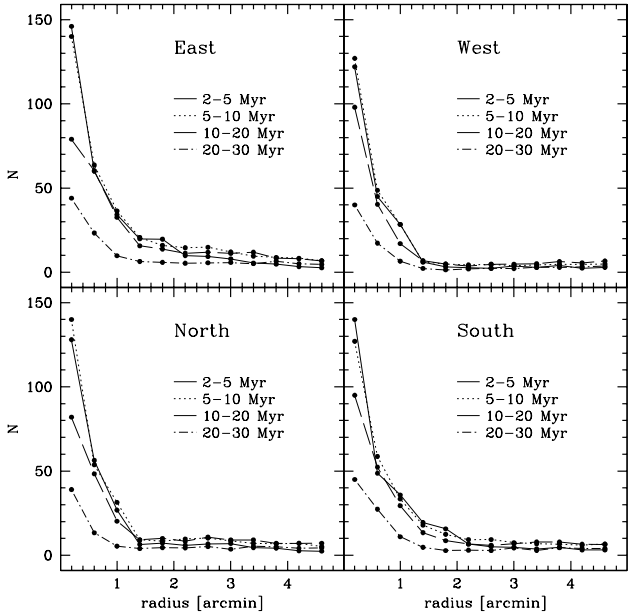}
\caption{Number of sources, per unit area, as a function of distance from the
centre of the cluster, in the various sectors of rings. North, South must be
intended respect to the Declination of the cluster centre, while East and West
respect to the Right Ascension of the centre. Symbols for the lines are the same
as in Fig.~\ref{profTOT}.}
\label{profSEP}
\end{figure}

With these data in hand, one can begin to determine more precisely the physical
structure of the cluster. To do this, we first plot the number of stars in each
age bin shown in Fig.~\ref{hist} inside annular rings 0.4' wide progressing in
steps of 0.4' from the cluster center out to a distance of 4.8' well outside the
nominal cluster radius. In Fig.~\ref{profTOT} we show this radial distribution
for the stars in each age bin. With these data we can confirm that the cluster
itself is totally contained within the reference radius of $2.5'$ obtained by
NPG02 since star number counts decrease until this value and remain constant at
larger radii. Stars in all the age ranges studied show the same behaviour,
indicating that all the PMS stars in our sample (old and young) definitely
belong to the cluster NGC~3603 as defined in Sect.~\ref{Intro}. Thus, all recent
star formation is confined to the compact cluster.\\  These results also show
that the objects located near the ZAMS in the CMD of Fig.~\ref{cmdiso} are not
field stars, but most of them belong to the cluster population. In fact, if we
assume that all the stars in the older bins, i.e. near the ZAMS, belong to the
field and not to the cluster and considering a uniform contamination in the
region, we would expect to obtain a flat profile, in contrast with what we
observe.  As expected, young stars (age $2 - 10$ Myr) in the considered mass
range are the most numerous ones, and they are concentrated towards the central
region of the cluster (see also Sect.~\ref{dens_map}). This result is similar to
that found in the massive cluster 30 Doradus, in the LMC \citep{demarchi11c} and
NGC~346 \citep[][in the Small Magellanic Cloud (SMC)]{demarchi11a,demarchi11b}. 
On the other hand, the cumulative radial distributions of young stars, defined
as all the PMS stars in our sample with ages between 2 and 10 Myr ($< 10$ Myr)
and old stars, all  PMS stars with ages between 10 and 30 Myr ($> 10$ Myr)
plotted in Fig.~\ref{cumPROF} yield an unexpected result. Although both curves
are steeper within $\sim 2.5'$ and tend to flatten beyond this radius, the
old/young ratio plotted in Fig.~\ref{cumRATIO} obtained from these two
distributions is almost flat, i.e. old stars are radially distributed roughly in
the same way as the young ones. This is further evidence supporting our previous
result that these stars are members of the cluster but it also shows that star
formation in NGC~3603 seems to have proceeded on average uniformly across the
cluster (but see the next Section for more details on this issue).\\
However, Fig.~\ref{profTOT} does not give us any information about the azimuthal
distribution because the number of objects is integrated over the whole area of
the rings. Therefore, to better characterize the structure of the cluster, we
divided each ring in two sectors of 180 degrees each, facing eastward and
westward  (with respect to the Right Ascension of the cluster centre) and
successively also  northward and southward (with respect to the Declination of
the cluster centre). We tried also to divide each ring in four sectors, but we
concluded that the number of objects would drop too much providing results with
poor statistics so we decided to keep the separation in the two halves described
above. The results of these star count profiles are shown in Fig.~\ref{profSEP},
where it is evident that the distribution of stars in the cluster is not
spherically symmetric. Actually, stars younger than 20 Myr seem more extended in
the South-East (SE) direction than in the North-West (NW) one.   

\subsection{Spatial distribution}
\label{dens_map}

\begin{figure*}
\centering
\hspace{-0.8cm}
\includegraphics[width=16cm]{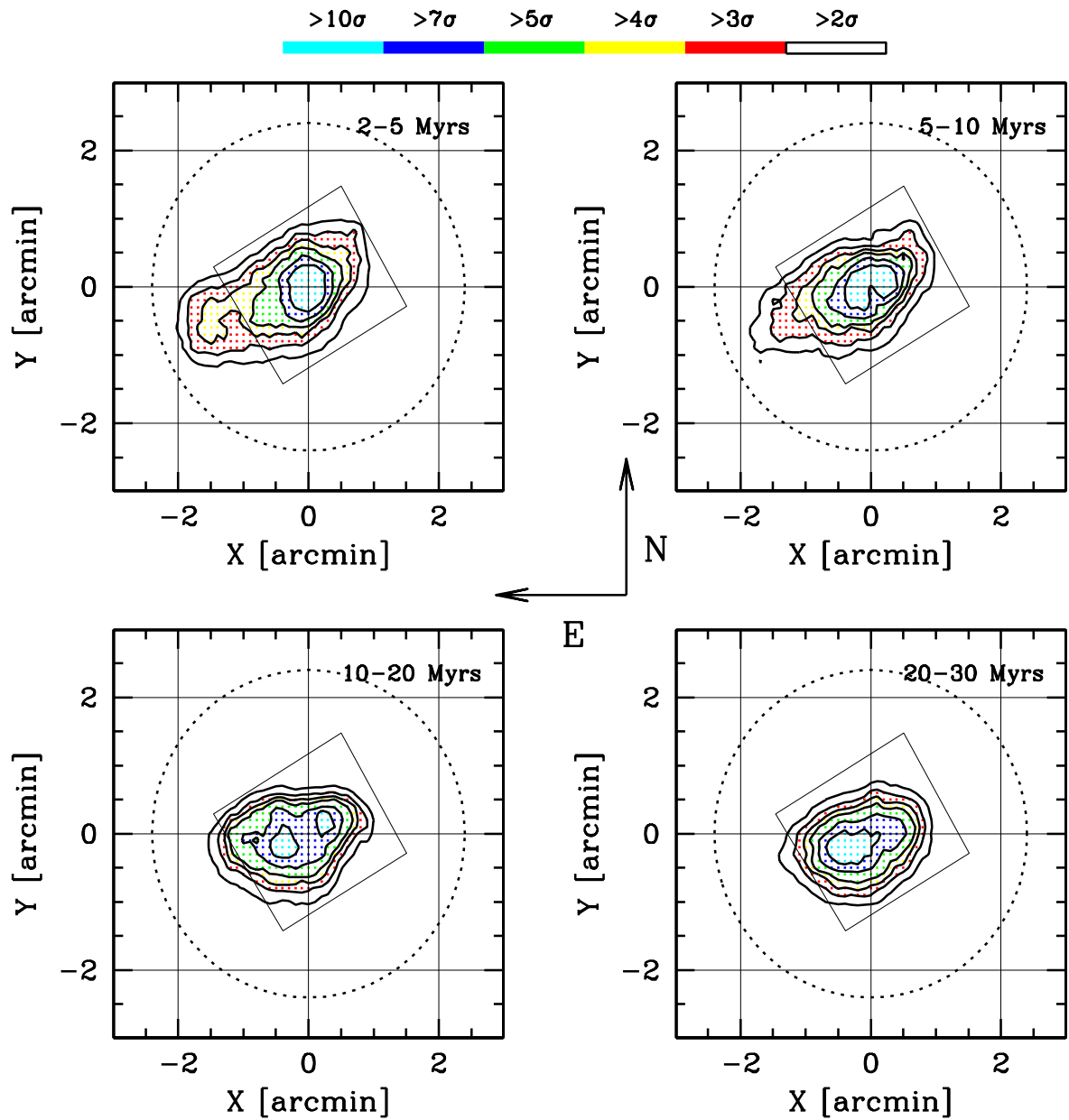}
\caption{Spatial distribution of stars in the same age intervals used for the
radial profiles (top left panel: 2-5 Myr; top right panel: 5-10 Myr; bottom
left panel: 10-20 Myr; bottom right panel: 20-30 Myr). Figure shows a zoomed
region of our FoV, with an area of $3'\times 3'$, centred on the centre of the
cluster determined by SB04. Colour code representing the number density of stars
above the background, expressed in terms of $\sigma$, is shown in the top part of 
the figure, the cyan colour indicating the density peak (i.e.: $>10\sigma$ over
the background); a circular area of radius $2.5'$, representing the nominal
radius of the cluster is plotted together with a square region representing the
WFC3 IR FoV. Orientation of the image is also reported.}
\label{map}
\end{figure*}

To better study and characterize the spatial distribution of the cluster and to
get a direct visual impression of the bi-dimensional shape and distribution of
the stars of different ages, we derived density maps obtained using the same
age  intervals as in the previous sections (see for example Fig.~\ref{profTOT}).
This allows us, in particular, to look for any possible differences in spatial
distribution between young and old stars.  To generate the density maps, we
divided the whole FoV in a grid, with a step of $0.1'$, and we counted stars
inside each bin. In this way, we derived a local number density for each point
across our FoV.  As a first step, we calculated a mean density for the
``background' analyzing only the external fields; in this way we obtained a mean
density, $\rho$, and a standard deviation, $\sigma$ that we used to derive the
final maps shown in Fig.~\ref{map}. In the density maps, we plotted with
different colours the regions where the density is $>3\sigma$, $>4\sigma$,
$>5\sigma$, $>7\sigma$ and $>10\sigma$ over the background for each age
interval, independently of the number of stars observed. We also show a contour
without colour, representing the region for which the density is $>2\sigma$, to
show that stars are really concentrated in the central region and do not extend
outside the nominal radius of $2.5'$.\\ In Fig.~\ref{map}, where we zoomed the
central region of the cluster, two aspects are easily observed: as already
stated before, stars preferentially extend in the SE direction, rather than in
the north or west ones. Moreover, it seems that the density peaks of young and
old stars are located in different positions in the field. Indeed, if we observe
the top left panel of Fig.~\ref{map}, we see that the centre of the cluster
coincides also with the density peak of stars with age between 2 and 5 Myr.  On
the other hand, moving towards older ages, we see in the top right panel of
Fig.~\ref{map} that the region of highest density seems to be more extended with
a curious shape with two lobes that encompass the centre of the cluster. Stars
between 10 and 20 Myr in the bottom left panel of Fig.~\ref{map} show two peaks
in the central region with the one located in  the SE more pronounced. Finally,
in the bottom right panel of Fig.~\ref{map} we find again a single peak of
density not spatially coincident with the centre of the cluster, but rather 
located towards the SE.\\ In conclusion, all the stars in the cluster are
distributed in a flattened oblate spheroidal pattern with the major axis
oriented in an approximate SE-NW direction and with a flattening or length of
the equatorial radius decreasing with increasing age. The density peak also
shifts towards the SE with increasing age. The origin of this peculiar asymmetry
is unclear but could be related to the possible  tidal effects on the cluster
stars by the massive gas clumps located above and below the cluster
\citep{nurn02b}. It may also be a reflection of the effects
of sequential triggering or positive feedback of the first older generation on 
the younger one due to winds and shocks compressing the molecular clouds 
surrounding the core. The substantial tail of young stars extending to the  SE
of the density peak towards the prominent pillar in that direction may be
tracing the region where the latter effect is most pronounced.\\ Curiously, the
most massive stars in the core of the cluster also seem to lie along a line
oriented in the same direction.

\section{Systematic uncertainties of the age determination}
\label{sys_unc}

Taken at face value, the results described so far clearly point to the cluster
stars in this mass range having a continuous age spread of at least $\sim 20-30$
Myr, with the star formation rate increasing continuously  up to at least
within $\sim 2$ Myr of the present time. However, this result  could be
seriously affected by the uncertainties listed in Sect.~\ref{Intro}. In this
section, these uncertainties are considered in detail and their effect measured
as accurately as possible to determine if they could alter the results described
above.

\subsection{Differential reddening}
\label{diff_redd}

\begin{figure}
\plotone{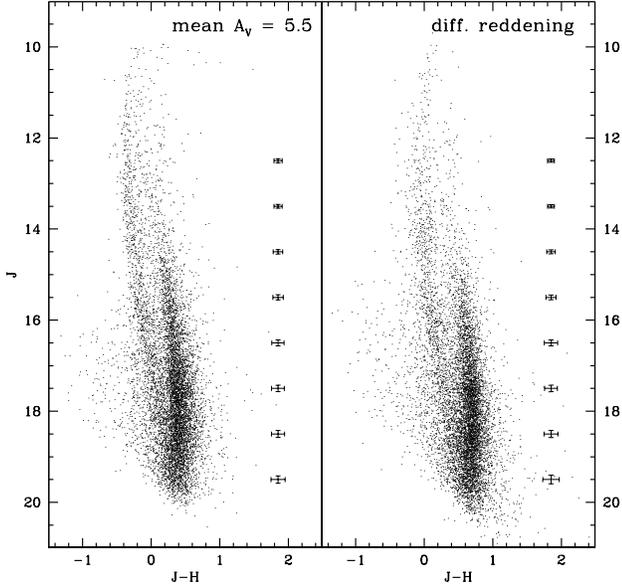}
\caption{Dereddened CMDs for the WFC3 field, obtained using the average total
extinction, $A_V = 5.5$ (left panel), and the correction for differential
reddening by P11 (right panel).}
\label{cfrebv}
\end{figure}

\begin{figure}
\plotone{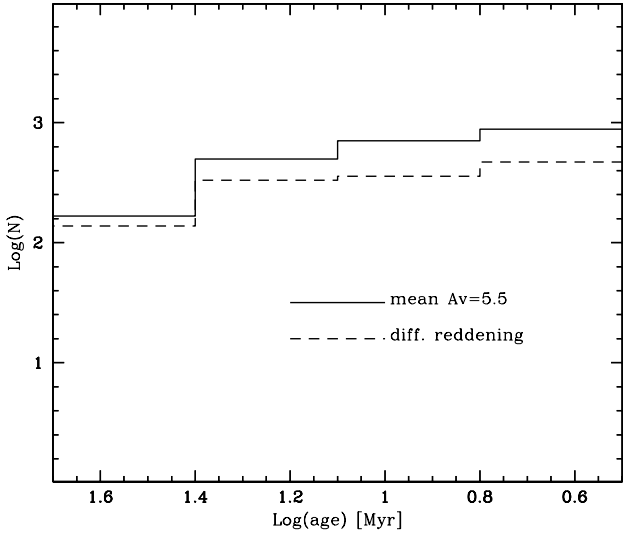}
\caption{Age histogram obtained using the mean $A_V = 5.5$ (solid line) compared
with the histogram obtained by star-to-star correction for differential
reddening from P11 (dashed line).}
\label{histebv}
\end{figure}

In our analysis so far, the possibly tried and true method of applying to all
the stars an average correction for reddening in an appropriately chosen wide
area was used, but this approach may lead to substantial errors if reddening
changes significantly on much smaller spatial scales. In order to take this
possible patchy or differential reddening into account, \citet[][hereafter
P11]{pang11} used the WFC3 H$\alpha$, Paschen$\beta$ filters and broad/medium
bands for the line continuum to compute the {\it Balmer} decrement of the gas
per pixel. By applying the \citet{fitz99} reddening law ($R_V = 3.1$), they
derived a two-dimensional map of colour-excess, $E(B-V)$ per pixel, across the
WFC3 FoV. Because the brighter stars in the core of NGC~3603 ($r < 0.52$ pc) are
saturated, P11 could not measure the colour excess associated with the core.
They adopted a fixed $E(B-V)=1.51$ mag, which is the median value of the colour
excess within an annulus ($0.52<r<1$ pc) centred on the core. This is justified
by the fact that, according to SB04 (see their Fig.~5), the colour excess within
1 pc from the core changes by only 0.05 mag.\\ As discussed in P11, there are
two sources of uncertainty in  this process: the first comes from the photon
noise of the images, and it varies on average from $\sigma_{E(B-V)} = 0.4$ at
$E(B-V) = 1.2$ mag to $\sigma_{E(B-V)} = 0.1$ mag at $E(B-V) = 2.4$ mag. The
second source of uncertainty is systematic and is due to the contribution of
emission lines to the flux in the broad/medium filters used to subtract the
continuum underlying the H$\alpha$ and Paschen$\beta$ lines. An estimated line
contamination of about 10\% in all broad/medium filters increases the colour
excess by 0.05 mag on average. The adopted extinction law also contributes to
the systematic uncertainties. By replacing the \citet{fitz99} extinction law
with that of \citet{cardelli89}, the colour excess becomes smaller by 0.1 mag.
On average, the systematic uncertainty is of the same order or smaller than the
uncertainty due to photon noise.\\ Here, we assume that the dust distribution in
NGC~3603 affects the gas and the stars in a similar way. Therefore, we adopt the
reddening map of P11 and read from it the  $E(B-V)$ at the spatial position of
each star detected in the $F110W$ and $F160W$ images. We assume the
\citet{cardelli89} extinction law with $R_V=3.55$ (from SB04) and correct for
extinction the $J$ and $H$ magnitudes of the detected stars. An uncertainty of
0.1 mag on the $E(B-V)$ derived by P11 translates into $\vartriangle (J-H) =
0.05$ mag. The result of this correction for the average plus the differential
reddening for the central WFC3 field is shown in the right panel of
Fig.~\ref{cfrebv}. For comparison, on the left panel of Fig.~\ref{cfrebv}, we
show the usual CMD obtained with the correction adopted so far ({\it i.e.}:
correction with the mean value $A_V=5.5$ for all the stars in the WFC3 FoV).\\ 
As is evident from the comparison between the two CMDs dereddened in different
ways, the correction for differential reddening (right panel of
Fig.~\ref{cfrebv}) does not have a dramatic impact on the overall distribution
of stars in the $J,J-H$ plane indicating a) that the local dust distribution is
relatively uniform, b) that most of the cluster stars lie behind the absorbing
dust and c) that applying an average total extinction to the outer HAWK-I fields
will not significantly affect the observed scatter of the stars there either
once the appropriate local value of $A_V$ is determined. The age histogram
obtained using star by star correction in the WFC3 field is shown in
Fig.~\ref{histebv} (dashed line), together with that using a mean value for
$A_V$ ($A_V=5.5$, solid line).\\ 
An error in the assumption that the $E(B-V)$ of the stars is the same as that of
the gas does not affect the observed stellar distribution or scatter in the CMD,
as it can only translate it slightly up or down in the direction almost parallel
to the isochrones, shown in Fig.~\ref{cmdiso}. This uncertainty in other words
would affect the mass determination, but only very marginally the age
determination \citep[see also][]{demarchi11c}. The same can be
said also for our  assumed value of $R_V$; the use of other extinction ratios
present in literature could only shift the whole CMD, but this collective effect
cannot erase the observed age spread. 

\subsection{Stars with possible IR excess}
\label{IRexcess}

\begin{figure}
\plotone{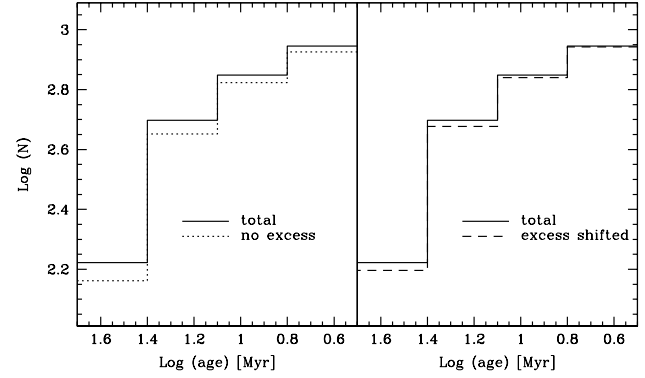}
\caption{Left panel: age histogram obtained removing all the sources with a
possible excess (dotted line) compared with the histogram with all the objects.
Right panel: age histogram obtained shifting the sources with an IR excess of a
mean quantity in $J$ and $H$, ($\delta<J> = 0.35$ and $\delta<H> = 0.69$,
respectively) derived by \citet[][,dashed line]{cieza05}. As in the
left panel, we show for comparison the usual histogram.}
\label{histPB}
\end{figure}

\begin{figure}
\plotone{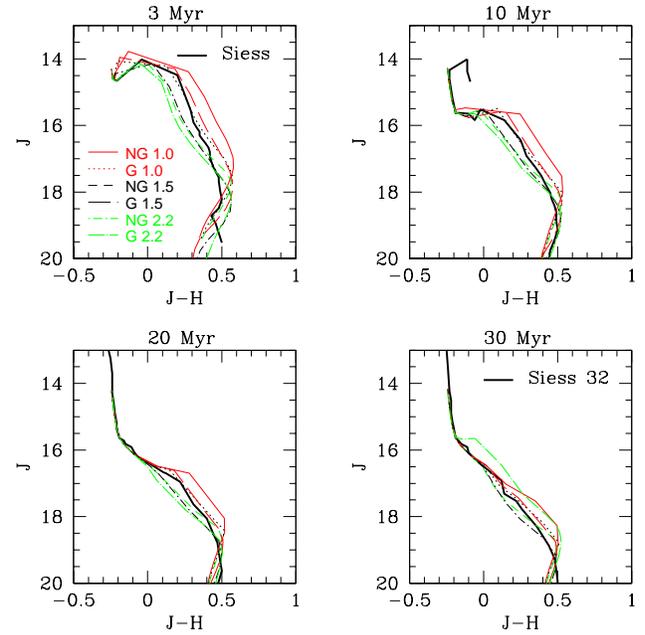}
\caption{The range of theoretical grey (G) and non grey atmosphere (NG)
isochrones from DC09, with convection efficiency represented by
$\alpha$ spanning from 1.0 to 2.2 plotted for 3, 10, 20, 30 Myr, together with
3, 10, 20, 32 Myr isochrones from \citet{siess00} for comparison.}
\label{compmod}
\end{figure}

\begin{figure}
\plotone{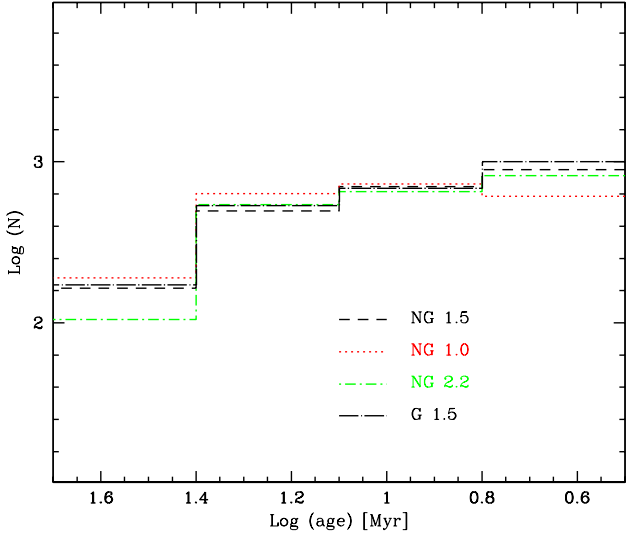}
\caption{Age histograms obtained using different theoretical models. The symbols
for the lines are the same as in Fig.~\ref{compmod}, so the black dashed line
represents the non grey atmosphere with $\alpha=1.5$, that is the model we adopted
in this paper. Red dotted line: non grey atmosphere with $\alpha=1.0$; green
dotted-short dashed line: non grey atmosphere with $\alpha=2.2$: black dotted-long
dashed line: grey atmosphere with $\alpha=1.5$. All models are consistent with
each other, except in the last bin, due to the slight differences among
isochrones (see Fig.~\ref{compmod})}.
\label{comphist}
\end{figure}
\begin{figure}
\plotone{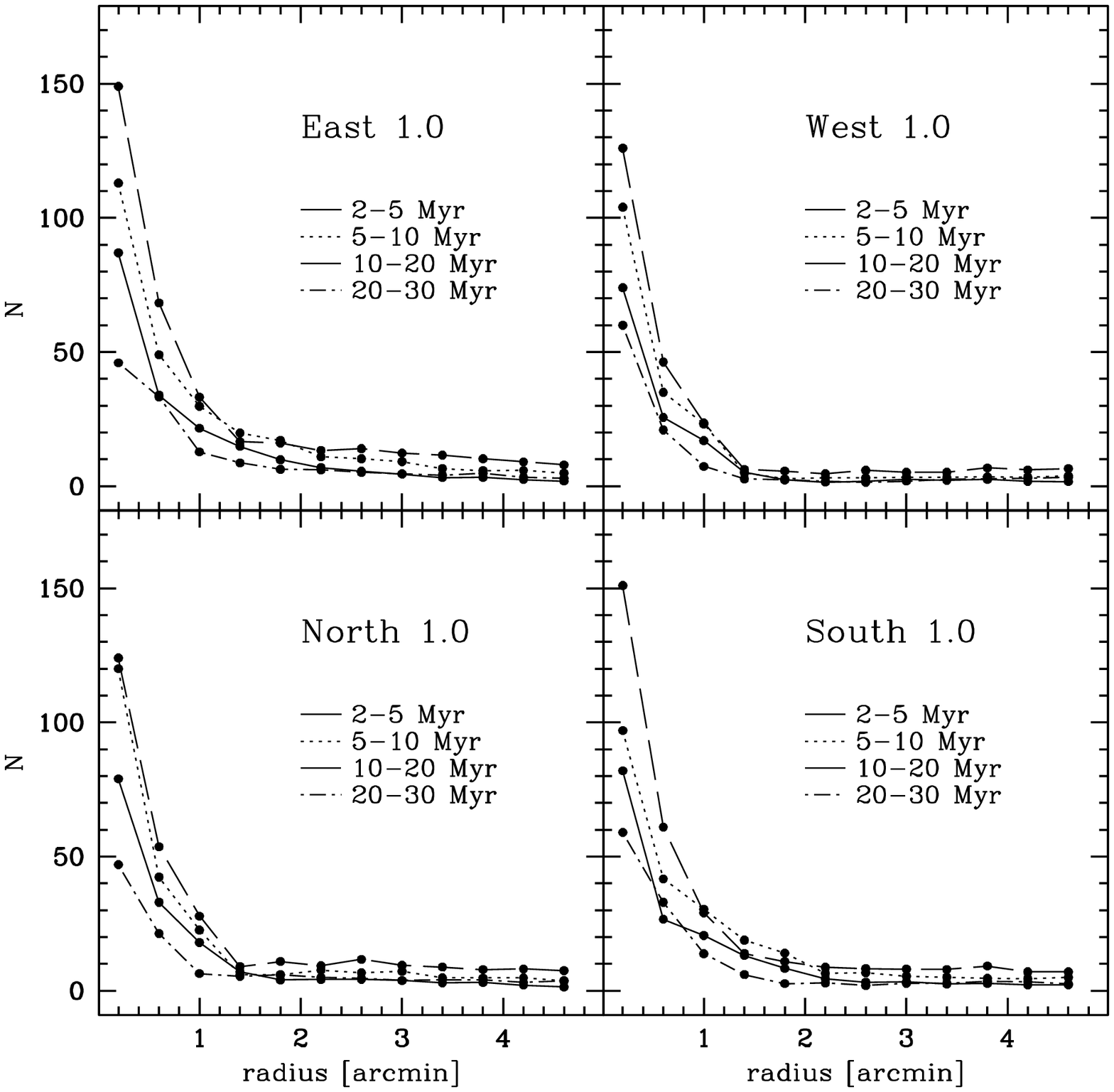}
\caption{Number of sources, per unit area, as a function of distance from the
centre of the cluster, in the various sectors of rings, using a theoretical
model with non grey atmosphere and $\alpha = 1.0$. North, South must be
intended respect to the Declination of the cluster centre, while East and West
respect to the Right Ascension of the centre. Symbols for the lines are the same
as in Fig.~\ref{profTOT}.}
\label{prof1}
\end{figure}
\begin{figure}
\plotone{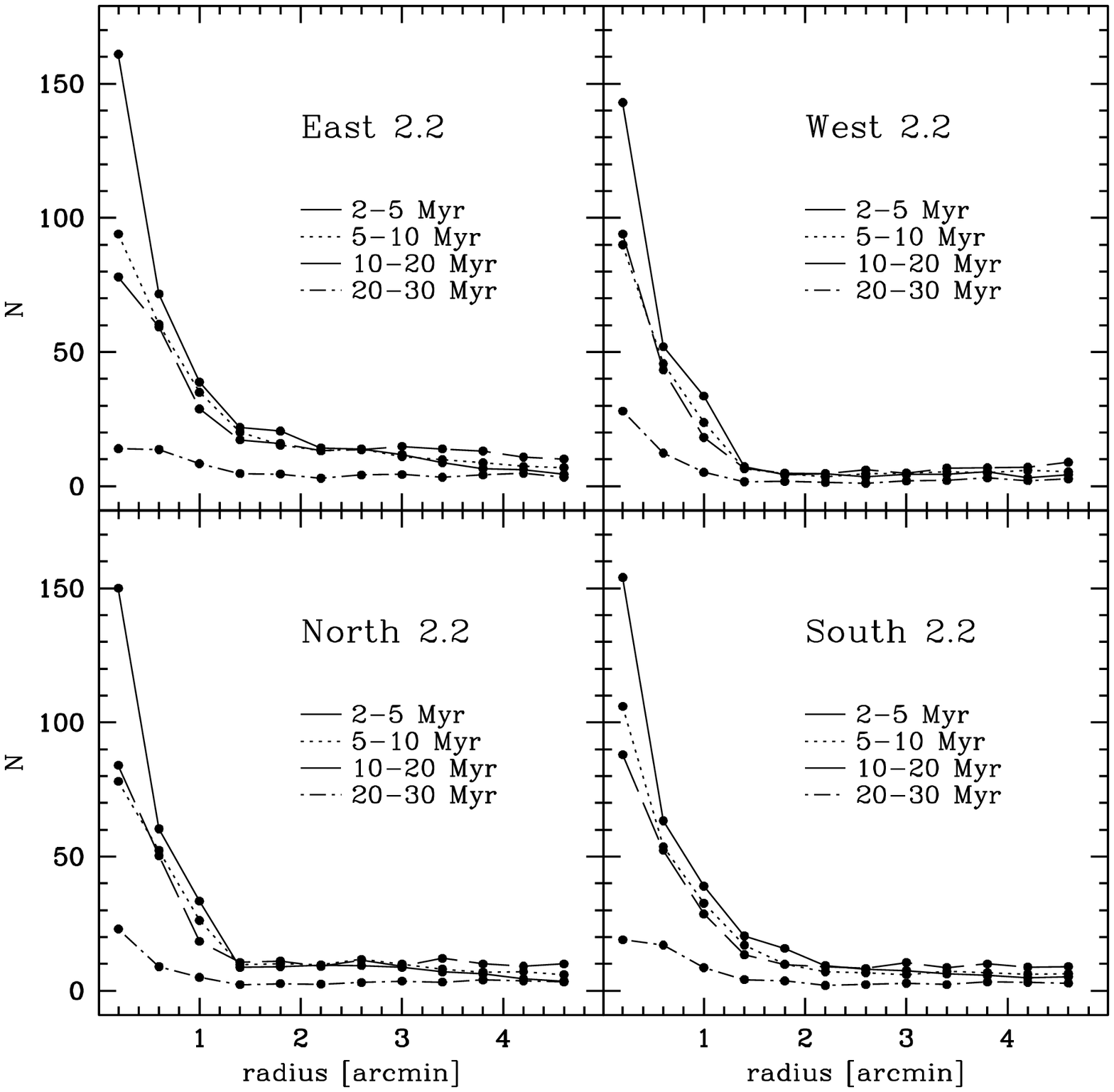}
\caption{Number of sources, per unit area, as a function of distance from the
centre of the cluster, in the various sectors of rings, using a theoretical
model with non grey atmosphere and $\alpha = 2.2$. North, South must be
intended respect to the Declination of the cluster centre, while East and West
respect to the Right Ascension of the centre. Symbols for the lines are the same
as in Fig.~\ref{profTOT}.}
\label{prof2}
\end{figure}
\begin{figure}
\plotone{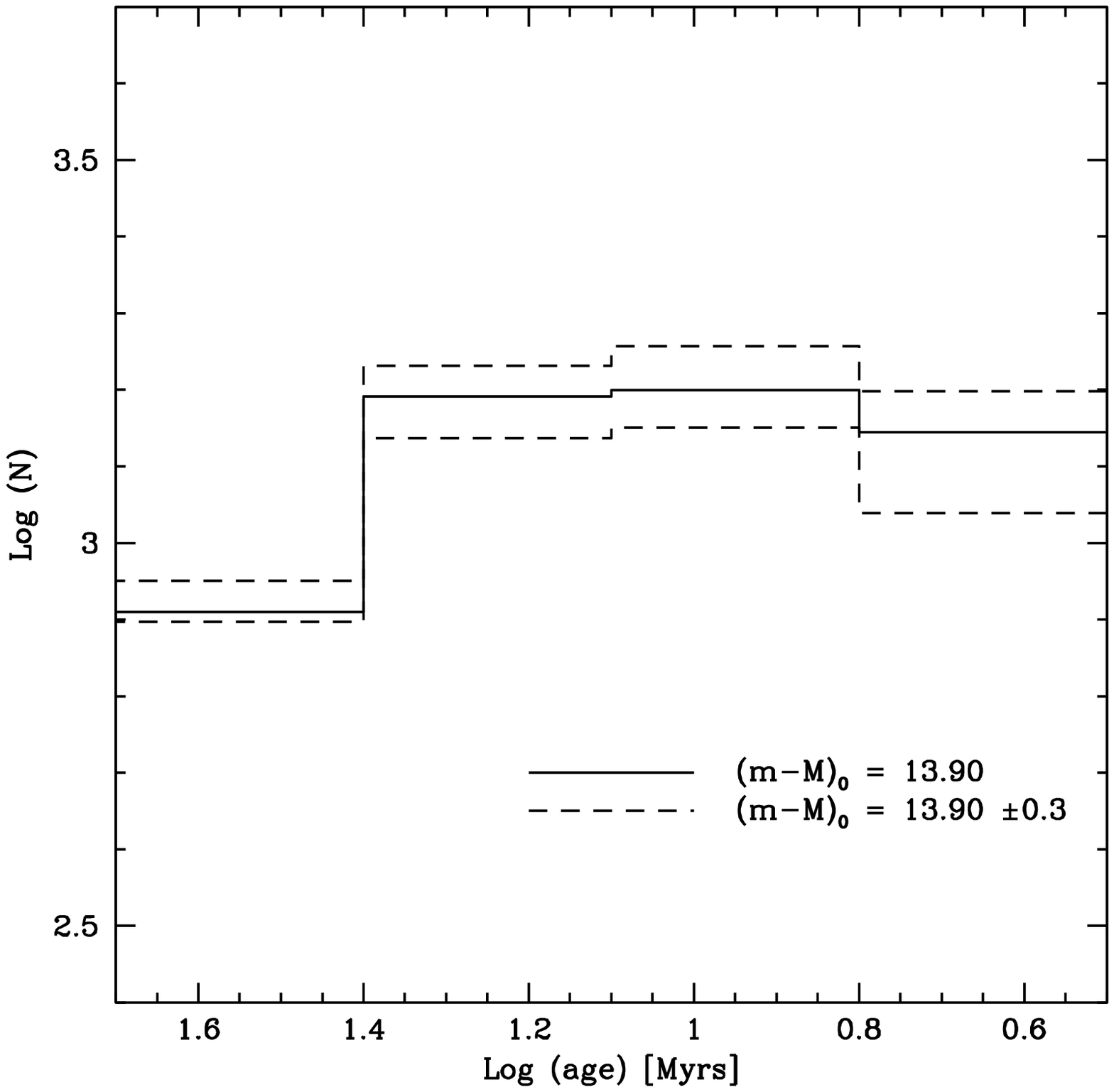}
\caption{Age histograms obtained using distance modulus of 13.90 mag (from
HEM08), compared with the two extreme cases for the uncertainty of this value
(distance modulus of 13.90 $\pm$ 0.3 mag, dashed lines).}
\label{histD}
\end{figure}

Another source of uncertainty could be an error in the correct location of stars
in the IR CMD due to reprocessing of photospheric emission by dust in a
circumstellar cloud or disc surrounding the object. This effect gives rise to an
excess that would need to be corrected for. A good handle on this effect can be
obtained by looking for indications of accretion activity in the PMS stars of
NGC~3603 by means of measurements of the H$\alpha$ and Paschen$\beta$ line
emission (contained in WFC3 F656N and F128N filters, respectively) in accreting
objects \citep[see for example][]{natta06}.\\ Among the 412 objects
with H$\alpha$ emission found in B10, 255 have an IR  counterpart. Exploiting
the F127M and F128N filters, and with a simple colour cut, $F127M-F128N \geq
0.35$, we found 802 objects with Paschen$\beta$ excess (a detailed description
of the selection criterion and of the properties of these stars is beyond the
scope of this paper and will be presented in a future work). Thus, the final
number  of stars showing an H$\alpha$ and/or Paschen$\beta$ excess, is 1057.
We removed them from the catalogue and we compared the age distribution obtained
with and  without them. We show this comparison in the left panel of
Fig.~\ref{histPB}, where the dashed line is the histogram between 2 and 30 Myr
obtained without them and the solid line is that showed in Fig.~\ref{histebv},
with all the objects and a mean $A_V$.\\  Another possibility we explored is to
quantify numerically in terms of magnitude the value of this excess and to shift
objects in the CMD by the appropriate amount. \citet{cieza05},
studying a sample of classical T Tauri stars, argued that they possess
significant non-photospheric excess in the $J$ and $H$ bands, and derived that
the mean excess in $J$ and $H$ is $\delta<J> \sim 0.35$ and $\delta<H> \sim
0.69$, respectively. The age distribution obtained after applying these
magnitude shifts to the 802 stars with Paschen$\beta$ excess is shown in the
right panel of Fig.~\ref{histPB}. This is compared as in the left panel with
the  histogram with all the objects. Looking at both cases, it is evident that
the trend observed is the same, either that we remove the stars or that we shift
them. Moreover,the comparison with the histogram with all the objects does not
show significant discrepancies. Hence, the three histograms are fully consistent
within the Poisson uncertainties. We can conclude that the source of uncertainty
due to IR excess in accreting stars is not a relevant problem in our work, and
cannot appreciably change the results reported here.

\subsection{Differences in theoretical models}
\label{diff_model}
In this paper we derive the ages of the stars of NGC~3603 using standard
hydrostatic stellar models. This procedure rests on the assumption that neither
the residual accretion after the protostellar phase nor the uncertainty in the
zero point of ages affect the results in a strong way. In any case, the
theoretical description of moderately low mass objects is affected by
uncertainties in the description of some physical inputs, in particular
convection and treatment of boundary conditions.\\ Since the low mass stars in
the PMS are fully convective and over-adiabatic, any change in the convective
model substantially alters the location of the track in the theoretical HR
plane. The use of a less efficient treatment of convection (lower $\alpha$)
leads to a larger temperature gradients, so that, for a given luminosity, the
structure readjusts on a more expanded configuration, with a subsequent shift of
the track to lower effective temperatures.\\ In order to see the effect of the
efficiency of convection on the age spread of NGC~3603 we computed a set of
models with three different values of the parameter
$\alpha=1,1.5,2.2$\footnote{According to \citet{dantona03}, the first value leads to a better agreement with the lithium
vs. T$_{eff}$ relation observed in young open clusters stars, while the last one
allows a fit of the solar radius for non-rotating models}. The path followed by
the theoretical PMS tracks on the HR diagram is also dependent on the boundary
conditions used to fit the numerical integration of the structural equations of
the interior with the atmosphere. In fact, the use of a non grey atmospheric
treatment shifts the tracks to cooler T$_{eff}$ within an extended interval of
masses and ages \citep{montalban04}. In order to show the
effect of grey models on the age of the stars, for
each choice of $\alpha$ we have also calculated models with a grey
approximation.\\  In Fig.~\ref{compmod} we compare the shape of isochrones of
different ages, in various models from DC09, which differ in the type of
atmosphere, grey or not grey, and in the value of the parameter of the
convection efficency $\alpha$. Moreover, we add to the comparison the model
from \citet[][adopted by B10 and S11]{siess00}.\\
The effect on the age histograms of the extreme cases is shown in
Fig.~\ref{comphist}, while in Fig.~\ref{prof1} and Fig.~\ref{prof2} we show what
happens to our radial profile distributions using the non grey atmosphere and
adopting $\alpha=1.0$ and $\alpha=2.2$.\\
All these models assume that protostars reach their birthline by constant or
slowly varying accretion. Recently, however, doubts have been expressed as to
the validity of this assumption when it was realized that if, instead,
protostars gain mass mainly by repeated but short lived episodes of disc
accretion, it may be possible for them to end up with much smaller radii and,
therefore, fainter than expected in the previous steady state accretion models.
This would have the effect of making them look much older than they really are
and biasing the cluster isochronal age distribution. Unfortunately, it is not
yet clear at all what this means in practice since the first analyses yield
contradicting results \citep{baraffe09,hosokawa11}.\\
Moreover, \citet{hartmann11} have argued persuasively that the cold episodic
accretion models used so far are unrealistic, in which case this scenario would
end up yelding uncertainties of only $\pm 1$ Myr in the age of low mass PMS
stars well within the measured age uncertainties of the older stars in our
sample. But until these issues are clarified by more theoretical and
observational work, movement of some young stars across the CMD due to episodic
accretion cannot be ruled out and even seems plausible, but it is quite unlikely
that it would affect such a large number of older stars in our sample. 

\subsection{Variability and distance uncertainties}
\label{var_dist}

Individual stars may vary in colour and luminosity somewhat of course, but there
is no indication from a comparison of our recent observations with the large
number of past ones that there have been any significant changes in the shape or
position of the PMS sequence in NGC 3603 that would alter in any way the
conclusion discussed here. The effect of uncertainties in the distance to NGC
3603 at a level of $\pm 0.3$ mag in the distance modulus (HEM08) translates into
a vertical displacement of $J=\pm 0.3$ mag in the IR CMD shown in
Fig.~\ref{cmdiso}. Although this small shift could in principle have an effect
as large as a factor of two on the absolute ages derived through comparison with
the isochrones, this would apply in a systematic way to all objects in the field
and therefore would not eliminate the age spread revealed by the CMD. This can
be easily observed in Fig.~\ref{histD} where we showed the usual histogram,
obtained with a distance modulus of 13.90 mag, compared with the two extremes in
the distance estimates  ($13.90 \pm 0.3$ mag, dashed lines). Finally, depth
effects are also negligible as the cluster diameter of 10 parsec represents at
most a depth variation of 10$^{-3}$ of the cluster distance.

\subsection{Unresolved binary systems}
\label{bin_stars}

Unresolved binaries can affect the uncertainties in the cluster age spread by
mimicking slightly brighter objects in the CMD and thereby being mistaken for
younger objects than observed. In other words, correction for unresolved
binaries would tend to flatten the age histogram shown in Fig.~\ref{hist} by
moving objects from the ``young'' to the ``old'' bins. A precise estimate of the
magnitude of this effect is difficult in the case of NGC~3603 because its binary
fraction and mass ratios have not been reliably measured so far but. 
Nonetheless, HEM08 have carried out a very detailed analysis of the effects that
unresolved binaries would have on the observed mass function of NGC~3603
obtained with ground based imaging. In their simulations, these authors assume
for NGC~3603 a binary fraction similar to that of the Orion Nebula Cluster,
owing to the the similar age and massive-star population of the two clusters,
and they rescale it accordingly for the different distance to NGC~3603.
Following HEM08, for the mass range of interest in our work ($\sim 1 -3$
$M_\odot$) we should assume a binary fraction of $\sim 5$\% and a mass ratio
somewhere between 0.5 and 1 (see HEM08 for details). Note that since the
limiting spatial resolution on our case is that of the HAWK-I photometry, the
direct comparison with the ground-based study of HEM08 is applicable. With
these values, the corresponding correction to the age histograms shown in
Fig.~\ref{comphist} for this effect would be far less than the error for each
bin due to the uncertainties in the theoretical models and, therefore, of no
practical consequence to the overall uncertainties in the measurement. A similar
result is obtained considering the probability of optical binaries namely
unresolved binaries due to projection effects that, with the resolution of the
WFC3, we estimate to be less than 1\% in the central WFC3 field with, therefore,
negligible consequence on the measured age spread.

\section{Summary and Conclusion}
\label{conclusion}

In this work, we set out to answer three basic questions concerning the stellar
population of the NGC~3603 cluster:
\begin{enumerate}
	\item Is the age spread discovered by B10 real or apparent?
	\item Is the age spread discovered by B10 cluster wide or confined to
	the central quarter covered by the WFC3 observations?
	\item And, if real, do the older stars actually belong to the cluster
	NGC~3603 or to the wider HII region surrounding it?
\end{enumerate}
With the data presented here, it seems reasonable
to assert that the age spread is indeed real and not due to other known sources 
of luminosity scatter in the CMD, and that it is cluster wide. In addition the 
older stars ($>10$ Myr) belong to the cluster and not to the surrounding field.
This means that, at least in the NGC~3603 cluster, star formation is relatively
slow, progressive and, to our limited temporal resolution, continuous. It
started some 20-30 Myr ago and extended almost to the present time uniformly
across the entire cluster. In some areas of the cluster (IRS~9, for example)
star formation is still ongoing probably because of triggering due to the impact
of the winds from the hot, massive stars in the core \citep{melena08}. 
An important corollary to our present study is that we find
good evidence for an asymmetric spatial distribution of the intermediate mass
cluster members varying with age that is most likely due to the by now well
established fact that star formation tends to occur in compact knots or cores of
gas and dust located along long filaments in the natal molecular cloud. This
hypothesis is made particularly plausible in our case since the large scale
distribution of gas and dust in the NGC 3603 GMC is elongated in a roughly NS
direction \citep{nurn02b} while on a smaller scale
around the star cluster, the orientation of the molecular material tends to lie
in the SE-NW direction roughly coinciding with that of the major axis of the
oblate spheroid. The reason for the observed variation with age in this case
would simply be due to the drift of the older stars away for their birhplace.  

\begin{acknowledgements}
M.C. and F.P. acknowledge the financial support of ASI through the ASI-MICELA AE
grant. We thank P. Grandi and P. Malaguti for useful discussions and suggestions
and for their support of the project and M. Mutchler for precious assistance in
data preparation and reduction. This paper is based on Early Release Science
observations made by the WFC3 Scientific Oversight Committee.  We are indebted
to the members of this committee for their support and encouragement in every
facet of our work. This publication makes use of data products from the Two
Micron All Sky  Survey, which is a joint project of the University of
Massachusetts and  Infrared Processing and Analysis Center/California Institute
of  Technology, funded by the National Aeronautics and Space Administration  and
the National Science Foundation. 
\end{acknowledgements}

%
\bibliographystyle{spr-mp-nameyear-cnd}  

\begin{thebibliography}{}

\bibitem[\protect\citeauthoryear{Allard \& Hauschildt}{1997}]{allard97}
Allard, F., \& Hauschildt, P.H., 1997, The NEXTGEN model grids. Web location:
\url{http://hobbes.hs.uni-hamburg.de/yeti/mdwarfs.html} B\"ohm-Vitense, E.
1958, Z. Astroph, 46, 1

\bibitem[\protect\citeauthoryear{Allard, Hauschildt \& Schweitzer}{2000}]{allard00}
Allard, F., Hauschildt, P.H., \& Schweitzer A., 2000, ApJ, 539, 366

\bibitem[\protect\citeauthoryear{Baraffe, Chabrier \& Gallardo}{2009}]{baraffe09}
Baraffe, I., Chabrier, G., \& Gallardo, J., 2009, ApJL, 702, L27

\bibitem[\protect\citeauthoryear{Beccari et al.}{2010}]{beccari10}
Beccari, G., et al., 2010, ApJ, 720, 1108 (B10)

\bibitem[\protect\citeauthoryear{Bellazzini et al.}{2002}]{bellazzini02}
Bellazzini, M., Fusi Pecci, F., Messineo, M., Monaco, L., \& Rood, R.~T., 2002,
AJ, 123, 1509

\bibitem[\protect\citeauthoryear{Bertin \& Arnouts}{1996}]{bertin96}
Bertin, E., \& Arnouts, S., 1996, A\&AS, 117, 393

\bibitem[\protect\citeauthoryear{Cardelli, Clayton \& Mathis}{1989}]{cardelli89}
Cardelli, J.A., Clayton, G.C., \& Mathis, J.S., 1989, ApJ, 345, 245 

\bibitem[\protect\citeauthoryear{Cieza et al.}{2005}]{cieza05}
Cieza, L.A., Kessler-Silacci, J.E., Jaffe, D.T., Harvey, P.M., \& Evans II, N.J.,
2005, AJ, 635, 422

\bibitem[\protect\citeauthoryear{D'Antona \& Montalb\`an}{2003}]{dantona03}
D'Antona, F., \& Montalb\`an, J., 2003, A\&A, 412, 213

\bibitem[\protect\citeauthoryear{De Marchi, Panagia \& Romaniello}{2010}]{demarchi10}
De Marchi, G., Panagia, N., \& Romaniello, M., 2010, ApJ, 715, 1  

\bibitem[\protect\citeauthoryear{De Marchi et al.}{2011a}]{demarchi11a}
De Marchi, G., Panagia, N., Romaniello, M., Sabbi, E., Sirianni, M., Prada 
Moroni, P.G., \& Degl'Innocenti, S., 2011a, ApJ, 740, 11

\bibitem[\protect\citeauthoryear{De Marchi, Panagia \& Sabbi}{2011b}]{demarchi11b}
De Marchi, G., Panagia, N., \& Sabbi, E., 2011b, ApJ, 740, 10

\bibitem[\protect\citeauthoryear{De Marchi et al.}{2011a}]{demarchi11c}
De Marchi, G., et al., 2011c, ApJ, 739, 27

\bibitem[\protect\citeauthoryear{De Pree, Nysewander \& Goss}{1999}]{depree99}
De Pree, C.G., Nysewander, M.C., \& Goss, W.M., 1999, AJ, 117, 2902

\bibitem[\protect\citeauthoryear{Di Criscienzo et al.}{2009}]{dicrisci09}
Di Criscienzo, M., Ventura, P., \& D'Antona, F., 2009, A\&A, 496, 223 (DC09)

\bibitem[\protect\citeauthoryear{Dotter}{2007}]{dotter07}
Dotter, A., 2007, PhDT, 17D

\bibitem[\protect\citeauthoryear{Eisenhauer et al.}{1998}]{eisenhauer98}
Eisenhauer, F., Quirrenbach, A., Zinnecker, H., \&  Genzel, R., 1998, ApJ, 498,
278

\bibitem[\protect\citeauthoryear{Fitzpatrick}{1999}]{fitz99}
Fitzpatrick, E.L., 1999, PASP, 111, 63

\bibitem[\protect\citeauthoryear{Grevesse \& Sauval}{1999}]{grevesse99}
Grevesse, N., \& Sauval, A.J., 1999, A\&A, 347, 348G

\bibitem[\protect\citeauthoryear{Harayama, Eisenhauer \& Martins}{2008}]{harayama08}
Harayama, Y., Eisenhauer, F., \& Martins, F., 2008, ApJ, 675, 1319 (HEM08)

\bibitem[\protect\citeauthoryear{Hartmann}{2003}]{hartmann03} 
Hartmann, L., 2003, ApJ, 585, 398

\bibitem[\protect\citeauthoryear{Hartmann, Zhu \& Calvet}{2011}]{hartmann11}
Hartmann, L., Zhu, Z., \& Calvet, N., 2011, arXiv1106.3343

\bibitem[\protect\citeauthoryear{Heiter et al.}{2002}]{heiter02}
Heiter, U., et al., 2002, A\&A, 392, 619

\bibitem[\protect\citeauthoryear{Hillenbrand}{2009}]{hillenbrand09} 
Hillenbrand, L.A., 2009, in The age of stars, Proceedings of the International
Astronomical Union, IAU Symposium, 258, 81

\bibitem[\protect\citeauthoryear{Hosokawa, Offner \& Krumholz}{2011}]{hosokawa11}
Hosokawa, T., Offner, S.S.R., \& Krumholz, M.R., 2011, ApJ, 738, 140

\bibitem[\protect\citeauthoryear{Jeffries}{2011}]{jeffries11} 
Jeffries, R.D., Proocedings of JENAM10: star clusters
in the era of large sueveys, in press, 2011, arXiv1102.4752J

\bibitem[\protect\citeauthoryear{Kalirai et al.}{2009}]{kalirai09}
Kalirai, J.S., et al., 2009, Instrument Science Report WFC3, 2009-30, 20 pages,
30

\bibitem[\protect\citeauthoryear{Kissler-Patig et al.}{2008}]{kissler08}
Kissler-Patig, M., et al., 2008, A\&A, 491, 941

\bibitem[\protect\citeauthoryear{Mackenty et al.}{2010}]{mackenty10}
Mackenty, J.W., Kimble, R.A., O'Connell, R.W., \& Townsend, J.A., 
2010, in Optical, Infrared and Millimeter Wave, edited by Oschmann J.M. Jr., 
Clampin M.C., MacEwen, H.A., Proocedings of the SPIE, vol. 7731, 77310

\bibitem[\protect\citeauthoryear{Melena et al.}{2008}]{melena08}
Melena, N.W., Massey, P., Morrell, N.I., \& Zangari, A.M., 2008, AJ, 135, 878

\bibitem[\protect\citeauthoryear{Melnick, Tapia \& Terlevich}{1989}]{melnick89}
Melnick, J., Tapia, M., \& Terlevich, R., 1989, A\&A, 213, 89 


\bibitem[\protect\citeauthoryear{Montalb\'an et al.}{2004}]{montalban04}
Montalb\'an, J., D'Antona, F., Kupka, F., \& Heiter, U., 2004, A\&A, 416, 1081

\bibitem[\protect\citeauthoryear{Natta, Testi \& Randich}{2006}]{natta06}
Natta, A., Testi, L., \& Randich, S., 2006, A\&A, 452, 245

\bibitem[\protect\citeauthoryear{N\"urnberger \& Petr-Gotzens}{2002}]{nurn02}
N\"urnberger, D.E.A., \& Petr-Gotzens, M.G., 2002, A\&A, 382, 537 (NPG02)

\bibitem[\protect\citeauthoryear{N\"urnberger et al.}{2002}]{nurn02b}
N\"urnberger, D.E.A., Bronfman, L., Yorke, H.W., \& Zinnecker, H., 2002, A\&A, 
394, 253 

\bibitem[\protect\citeauthoryear{Pang, Pasquali \& Grebel}{2011}]{pang11}
Pang, X., Pasquali, A., \& Grebel, E.K., 2011, AJ, 142, 132 (P11)

\bibitem[\protect\citeauthoryear{Preibisch et al.}{2011}]{preibisch11}
Preibisch, T., et al., 2011, A\&A, 530, A34

\bibitem[\protect\citeauthoryear{Rochau et al.}{2010}]{rochau10}
Rochau, B, Brandner, W., Stolte, A., Gennaro, M., Gouliermis, D., Da Rio, N., 
Dzyurkevich, N., \& Henning, T., 2010, ApJL, 716, L90

\bibitem[\protect\citeauthoryear{Siess, Dufour \& Fiorentini}{2000}]{siess00}
Siess, L., Dufour, E., \& Forestini, M., 2000, A\&A, 358, 593

\bibitem[\protect\citeauthoryear{Skrutskie et al.}{2006}]{skrutskie06}
Skrutskie, M.F., et al., 2006, AJ, 131, 1163 

\bibitem[\protect\citeauthoryear{Spezzi et al.}{2011}]{spezzi11}
Spezzi, L., et al., 2011, ApJ, 731, 1 (S11)

\bibitem[\protect\citeauthoryear{Stetson}{1987}]{stetson87}
Stetson, P.B., 1987, PASP, 99, 191

\bibitem[\protect\citeauthoryear{Stetson}{1994}]{stetson94}
Stetson, P.B., 1994, PASP, 106 ,250

\bibitem[\protect\citeauthoryear{Stolte et al.}{2004}]{stolte04}
Stolte, A., Brandner, W., Brandl, B., Zinnecker, H., \& Grebel, E.K., 2004, AJ,
128, 765

\bibitem[\protect\citeauthoryear{Sung \& Bessell}{2004}]{sung04}
Sung, H., \& Bessell, M.S., 2004, AJ, 127, 1014 (SB04)

\bibitem[\protect\citeauthoryear{Ventura, D'Antona \& Mazzitelli}{2008}]{ventura08}
Ventura, P., D'Antona, F., \& Mazzitelli, I., 2008, Ap\&SS, 316, 93

\bibitem[\protect\citeauthoryear{Wong et al.}{2010}]{wong10}
Wong, M.H., et al., 2010, Wide Field Camera 3 Instrument  Handbook, Version 2.0,
Baltimore: STSCI

\end{thebibliography}

\appendix

\section{Description of isochrones}
\label{Appendix}

The isochrones are determined by interpolating among evolutionary sequences of
models of different mass, calculated by means of the ATON code for stellar
evolution \citep{ventura08}. The evolutions are followed from a
fully convective configuration, with initial central temperatures $T_C$ in the
range $5.3 < logT_C < 5.8$, until the beginning of the core H-burning phase.
Both the deuterium and lithium PMS burning are followed.\\ The chemistry used is
$(X,Y,Z) = (0.70,0.28,0.02)$; the distribution of the heavy elements follows
the solar mixture by \citet{grevesse99}. The initial mass
fraction of Deuterium is $X(D)=2\times10^5$. These models are based on up to
date physics and updated non grey atmosphere models were used based on the
NextGen models by \citet{allard97}, complemented by the low
gravity models by \citet{allard00}. Following a suggestion by
\citet{heiter02}, we match the atmospheric grid with the interior
integration at $\tau = 10$. This choice should minimize the consistency problems
related to the different EOS and opacities adopted in the interior and in the
atmosphere, and to the absence of turbolence pressure in the atmospheric
modeling.\\ The temperature gradient within regions unstable to convection is
found by the Mixing Length Theory, with the free parameter $\alpha = 1.5$. This
choice for $\alpha$ accounts for the smaller efficiency of convection in the PMS
phase, as indicated by the location in the HR diagram of the few objects for
which the mass is known, and by the Lithium vs Mass pattern detected in young
open clusters \citep{dantona03}. The L and $T_{eff}$
values, are then transformed into the different observational magnitudes using
the synthetic colour transformations kindly provided by \citep[see][and references
therein]{dotter07}.

\end{document}